\documentclass[showpacs,showkeys,preprintnumbers,amsmath,amssymb,latexsym,floatfix,superscriptaddress,twocolumn,prd]{revtex4}

\usepackage{graphicx}
\usepackage{dcolumn}
\usepackage{bm}

\makeatletter    

\def\meqalign#1{\null\,\vcenter{\openup\jot\m@th \let\\=\crcr 
  \ialign{\strut\hfil$\displaystyle{##}$&&$\displaystyle{{}##}$\hfil
      \crcr#1\crcr}}\,}

\makeatother   

\begin{document}

\title{$e^--e^+$ pair creation by vacuum polarization around electromagnetic black holes}

\author{C. Cherubini}
 \affiliation{ICRANet, I-65100 Pescara, Italy}
  \affiliation{Nonlinear Physics and Mathematical Modeling Lab, C.I.R., University Campus Bio-Medico, I-00128 Rome, Italy
}
\author{A. Geralico}
 \affiliation{ICRANet, I-65100 Pescara, Italy}
  \affiliation{Physics Department and
ICRA, University of Rome ``La Sapienza,'' I-00185 Rome, Italy
}
\author{J. A. Rueda H.}
 \affiliation{ICRANet, I-65100 Pescara, Italy}
  \affiliation{Physics Department and
ICRA, University of Rome ``La Sapienza,'' I-00185 Rome, Italy
}
\author{R. Ruffini}
 \email{ruffini@icra.it}
 \affiliation{ICRANet, I-65100 Pescara, Italy}
  \affiliation{Physics Department and
ICRA, University of Rome ``La Sapienza,'' I-00185 Rome, Italy
}

\date{\today}

\begin{abstract}
The concept of ``dyadotorus'' was recently introduced to identify in the Kerr-Newman geometry the region where vacuum polarization
processes may occur, leading to the creation of $e^--e^+$ pairs. This
concept generalizes the original
concept of ``dyadosphere'' initially introduced for Reissner-Nordstr\"{o}m
geometries.  The topology of the axially symmetric dyadotorus is studied for
selected values of the electric field and its electromagnetic energy is estimated by using three different methods all giving the same result.
It is shown by a specific example the difference between a dyadotorus and a dyadosphere.
The comparison is made for a Kerr-Newman black hole with the same total mass energy and the same charge to mass
ratio of a Reissner-Nordstr\"{o}m black hole. It turns out that the Kerr-Newman black hole leads to larger values
of the electromagnetic field and energy when compared to the electric field and
energy of the Reissner-Nordstr\"{o}m one.
The significance of these theoretical results for the realistic description of the process of gravitational collapse leading to black hole formation as well as the energy source of gamma ray bursts are also discussed.
\end{abstract}



\maketitle

\section{Introduction}

Relativistic astrophysics differs from the other branches of physics and astronomy by exploring new fundamental processes unprecedented for the spectacularly large scales of the involved observables and for their extremely short time variability.
Following the well known case of supernova with energies $\lesssim10^{53}$ ergs on time scales of months, gamma-ray bursts (GRBs) have offered an extreme example of the most energetic ($E\lesssim10^{55}$ ergs) and the fastest transient ($\Delta t\lesssim10^{-3}-10^4$ s) phenomena ever observed in the universe \cite{brasile}.
The dynamics of GRBs is dominated by an electron-positron plasma \cite{report}.
The theoretical model based on the vacuum polarization processes \cite{damruff} occurring in a Kerr-Newman geometry \cite{ruffinikerr} can indeed explain such enormous energetics and the sharp time variability.
What is most important is that such a model is based on explicit analytic solutions of well-tested ultrarelativistic field theories.
The formation of such black holes in a process of gravitational collapse is expected from a large variety of binary mergers composed of neutron stars, white dwarfs and massive stars at the end point of their thermonuclear evolution \cite{ruffinimg11} in all possible combinations.

In particular, in the merging of two neutron stars and in the final process
of gravitational collapse to a black hole is expected the occurrence of electromagnetic fields
with strength larger than the critical value of vacuum polarization
\begin{equation}
\label{Ecrit}
E_c=\frac{m_e^2c^3}{\hbar e}\ ,
\end{equation}
where $m_e$ and $e$ are the electron mass and charge respectively
\cite{brasile}. We are currently reexamining the electrodynamical processes of
a neutron star via an ultrarelativistic Thomas-Fermi equation to identify the
possible physical origin of this overcritical electric field
\cite{rotondoijmp,patricelli,rueda}.

The time evolution of the gravitational collapse (occurring on characteristic gravitational time scales $\tau=GM/c^3\simeq 5\times 10^{-5} M/M_\odot$ s) and the associated electrodynamical process  are too complex
for a direct description.  We address here a more confined problem: the polarization process around an already formed Kerr-Newman black hole.
This is a well defined theoretical problem which deserves attention. It represents a physical state asymptotically reachable in the process of gravitational collapse. We expect such an asymptotic configuration be reached when all
the multipoles departing from the Kerr-Newman geometry have been radiated away either by process of vacuum polarization
or electromagnetic and gravitational waves.
What it is most important is that by performing this theoretical analysis we can gain a direct evaluation
of the energetics of the spectra and dynamics of the $e^--e^+$ plasma created on the extremely short time scales due to the quantum phenomena of $\Delta t=\hbar/(m_e c^2)\simeq 10^{-21}$ s.
This entire transient phenomena, starting from an initial neutral condition, undergoes the formation of the Kerr-Newman black hole by the collective effects of gravitation, strong, weak, electromagnetic interactions during a fraction of the above mentioned gravitational characteristic time scale of collapse.

The aim of this article is to explore the initial condition for such a process to occur
using the recently introduced concept of ``dyadotorus'' \cite{ruffinikerr} which generalizes to the
Kerr-Newman geometry the concept of the ``dyadosphere'' previously introduced in the case of the
spherically symmetric Reissner-Nordstr\"om geometry \cite{rr1,prx}.

Damour and Ruffini \cite{damruff} showed that vacuum polarization processes
{\it \`{a} la} Sauter-Heisenberg-Euler-Schwinger
\cite{sauter,heiseul,schwinger} can occur in the field of a Kerr-Newman black
hole endowed with a mass ranging from the maximum critical mass for neutron
stars $(3.2M_{\odot})$ all the way up to $7.2\times10^6M_{\odot}$.
It is an almost perfectly reversible process in the sense defined by Christodoulou and
Ruffini \cite{chrruff}, leading to a very efficient mechanism of extracting
energy from the black hole.

In the case of absence of rotation in spacetime, we have a
Reissner-Nordstr\"{o}m black hole as the background geometry. The region where
vacuum polarization processes take place is a sphere centered about the hole,
and has been called dyadosphere \cite{rr1,prx}. Its main properties are
recalled in Section II.

We investigate in Section III how the presence of rotation in spacetime
modifies the shape of the surface containing the region where electron-positron
pairs are created.
Due to the axial symmetry we call that region as dyadotorus and we give the conditions for its existence.
We then provide some pictorial representations of the boundary surface of the dyadotorus by
using the Boyer-Lindquist radial and angular coordinates as polar coordinates
in flat space as well as by employing Kerr-Schild coordinates. We show in
Section IV the dyadotorus on the corresponding embedding diagrams, which reveal
the intrinsic structure of the spacetime geometry. In Section V we provide an
estimate of the electromagnetic energy contained in the dyadotorus by using three
different definitions commonly adopted in the literature, i.e. the standard
definition in terms of the timelike Killing vector (see e.g. \cite{vitagliano}), the one recently
suggested by Katz, Lynden-Bell and Bi{\v c}\'ak \cite{katz1,katz2} for axially
symmetric asymptotically flat spacetimes, which is an observer dependent definition of energy, and the
last one involving the theory of pseudotensors (see e.g. \cite{virbhadra}). All these approaches are shown to give
the same results. Finally, a comparison is made between the electromagnetic energy of an extreme Kerr-Newman black hole
and the corresponding one of a Reissner-Nordstr\"om black hole with the same total mass and charge to
mass ratio. In addition to the topological differences between the dyadotorus and the dyadosphere, it is shown how
larger field strengths are allowed in the case of a Kerr-Newman geometry close to the horizon, when compared with a
Reissner-Nordstr\"om black hole of the same mass energy and charge to mass ratio.

We finally draw some general conclusions.

\section{The dyadosphere}

In this section we recall the definition of dyadosphere and its main properties
in the field of a Reissner-Nordstr\"{o}m black hole as derived in
\cite{rr1,prx}. In standard Schwarzschild-like coordinates the
Reissner-Nordstr\"om black hole metric is given by
\begin{eqnarray}
\label{RNmetric}
\hspace{-.1cm}ds^2\hspace{-.1cm}&=&\hspace{-.1cm}- \left(1 - \frac {2M}{r}+\frac{Q^2}{r^2}\right)dt^2 + \left(1 - \frac {2M}{r}+\frac{Q^2}{r^2}\right)^{-1}dr^2\nonumber\\
&&+r^2(d\theta^2 +\sin ^2\theta d\phi^2)\ ,
\end{eqnarray}
where geometric units $G=c=1$ have been adopted.  The associated
electromagnetic field is given by
\begin{equation}
\label{RNemfield}
F=-\frac{Q}{r^2}dt\wedge dr\ .
\end{equation}
The horizons are located at $r_\pm=M\pm\sqrt{M^2-Q^2}$;
we consider the case $|Q|\leq M$ and the region $r>r_+$ outside
the outer horizon.  For an extremely charged hole we have $|Q|=M$
and the two horizons coalesce.

Let us introduce an orthonormal frame adapted to the static observers
\begin{eqnarray}
\label{frameRN}
e_{\hat t}&=&\left(1 - \frac {2M}{r}+\frac{Q^2}{r^2}\right)^{-1/2}\partial_t\ , \nonumber\\
e_{\hat r}&=&\left(1 - \frac {2M}{r}+\frac{Q^2}{r^2}\right)^{1/2}\partial_r\ , \nonumber\\
e_{\hat \theta}&=&\frac{1}{r}\,\partial_\theta\ , \qquad
e_{\hat \phi}=\frac{1}{r\sin \theta}\,\partial_\phi\ .
\end{eqnarray}
The electric field as measured by static observers with four-velocity $U=e_{\hat t}$ is purely radial
\begin{equation}
E(U)=\frac{Q}{r^2}\,e_{\hat r}\ .
\end{equation}
The radius $r_{\rm ds}$ at which the electric field strength $|{\bf
E}|=|E^{\hat r}|$ reaches the critical value $E_c$ has been defined in
\cite{rr1,prx} as the outer radius of the dyadosphere, which extends down to
the horizon and within which the electric field strength exceeds the critical
value
\begin{equation}
r_{\rm ds}\simeq1.12\times10^8\sqrt{\lambda\mu}\,\,\mbox{cm}\ ,
\end{equation}
where the dimensionless quantities $\lambda=Q/M$ and $\mu=M/M_\odot$ have been introduced.
The critical electric field (\ref{Ecrit}) in geometric units is given by $E_c\approx1.268\times10^{-11}$ cm$^{-1}$.

\begin{figure}
\typeout{*** EPS figure 0a}
\begin{center}
\includegraphics[scale=0.45]{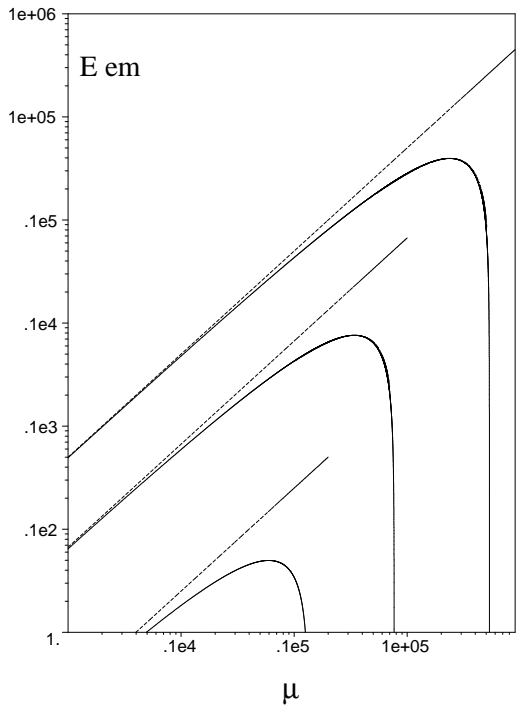}
\caption{The behavior of the electromagnetic energy (\ref{enevita}) in solar
mass units is shown as a function of the mass parameter $\mu$ for selected
values of the charge parameter $\lambda=[0.1,0.5,1]$, from bottom to top. The
straight lines (dashed) correspond to the maximum energy extractable from a
Reissner-Nordstr\"om black hole given by $Q^2/(2r_+)$.} \label{fig:0a}
\end{center}
\end{figure}

The electromagnetic energy contained inside the dyadosphere has been evaluated
by Vitagliano and Ruffini \cite{vitagliano}
\begin{equation}
\label{enevita}
E(\xi)_{(r_+,r_{\rm ds})}=\frac{Q^2}{2r_+}\left(1-\frac{r_+}{r_{\rm ds}}\right)\ ,
\end{equation}
by using a ``truncated version'' of the definition of energy in terms of the
conserved Killing integral
\begin{equation}
E(\xi)=\int_{\Sigma} T_{\mu\nu}^{(\rm em)}\xi^\mu d\Sigma^\nu\ ,
\end{equation}
where $\xi=\partial_t$ is the timelike Killing vector.
We refer to Section V for a detailed discussion on this point. Fig.
\ref{fig:0a} shows the behaviour of the electromagnetic energy (\ref{enevita})
as a function of the mass parameter $\mu$ for fixed values of the charge
parameter $\lambda$.


\begin{figure}
\typeout{*** EPS figure 0b}
\begin{center}
\includegraphics[scale=1.2]{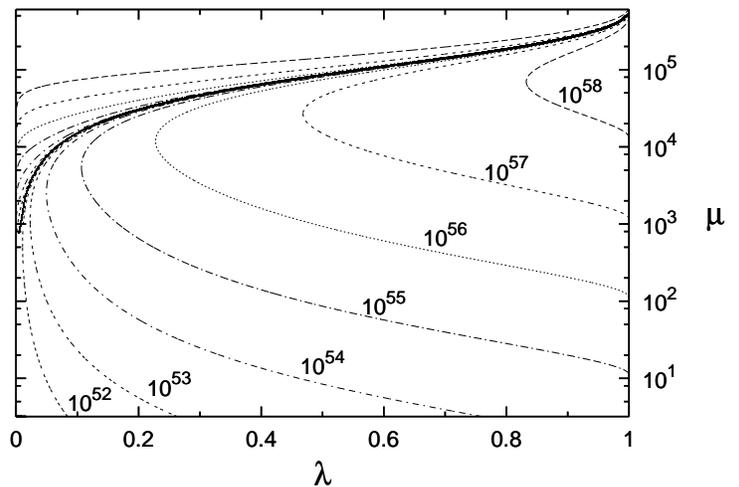}
\caption{The total energy of pairs (\ref{Epairs}) is plotted as a function of
the two mass and charge parameters $\mu$ and $\lambda$. The different curves
correspond to selected values of the the energy (in ergs). Only the solutions
below the solid line are physically relevant. The configurations above the
solid line correspond instead to unphysical solutions with $r_{\rm ds}<r_+$. 
The plot is reproduced from \cite{brasile2} with the kind permission of the authors.}
\label{fig:0b}
\end{center}
\end{figure}

Ruffini and collaborators estimated also the total energy of pairs converted
from the \lq\lq static electric energy'' (\ref{enevita}) and deposited within
the dyadosphere
\begin{equation}
\label{Epairs}
E_{\rm pairs}=\frac{Q^2}{2r_+}\left(1-\frac{r_+}{r_{\rm ds}}\right)\left[1-\left(\frac{r_+}{r_{\rm ds}}\right)^4\right]\ .
\end{equation}
Its behaviour as a function of the charge and mass parameters $\lambda$ and $\mu$
is shown in Fig. \ref{fig:0b}.


The rate of pair creation per unit four-volume is given by the Schwinger formula \cite{schwinger}
\begin{equation}
\label{rateRN}
2{\rm Im}\mathcal L=\frac1{4\pi}\left(\frac{|{\bf E}|e}{\pi\hbar}\right)^2\sum_{n=1}^\infty \frac{1}{n^2}e^{-n\pi E_c/|{\bf E}|}\ .
\end{equation}
The leading term $n=1$ agrees with the WKB results obtained by Sauter \cite{sauter} and Heisenberg-Euler \cite{heiseul}
\begin{equation}
2{\rm Im}\mathcal L=\frac1{4\pi}\left(\frac{|{\bf E}|e}{\pi\hbar}\right)^2e^{-\pi E_c/|{\bf E}|}\ .
\end{equation}

The dyadosphere has been defined by Ruffini and collaborators \cite{rr1,prx} by the condition $|{\bf E}|=E_c$.
One might better define it by requiring the electric field strength to be such
that the rate of pair creation is suppressed exactly by a factor $1/e$, leading
to the condition $|{\bf E}|=\pi E_c$. However, from Eq. (\ref{rateRN}) it is
clear that no sharp threshold exists for electron-positron pair creation, so
that the definition
\begin{equation}
\label{dyadoredef}
|{\bf E}|=k E_c\
\end{equation}
appears to be more appropriate and should be explored for different values
of the constant parameter $k$, even for $k<1$.  Consequently, we shall define
in the following both dyadosphere and dyadotorus as the locus of points where the
electric field satisfies the condition (\ref{dyadoredef}).

\section{The dyadotorus}

The Kerr-Newman metric in standard Boyer-Linquist type coordinates writes as \cite{MTW}
\begin{eqnarray}
\label{KNmetric}
\hspace{-1cm}ds^2\hspace{-.1cm}&=&\hspace{-.1cm}-\left( 1- \frac{2Mr-Q^2}{\Sigma} \right) dt^2\nonumber\\
&&\hspace{-.1cm} -\frac{ 2a\sin^2\theta }{\Sigma}\left(2Mr-Q^2\right) dt d\phi+\frac{\Sigma}{\Delta} dr^2+\Sigma d\theta^2\nonumber\\
&&\hspace{-.1cm}+ \left[r^2+a^2+\frac{a^2\sin^2\theta}{\Sigma}(2Mr-Q^2)\right]\sin^2\theta d\phi^2,
\end{eqnarray}
with associated electromagnetic field
\begin{eqnarray}
\label{Fkn}
F&=& \frac{Q}{\Sigma^2}(r^2-a^2\cos^2\theta)dr \wedge [dt -a \sin^2 \theta d\phi]\nonumber \\
&& +2\frac{Q}{\Sigma^2}ar \sin \theta \cos \theta d\theta \wedge [(r^2+a^2)d\phi - a dt],
\end{eqnarray}
where $\Sigma=r^2+a^2\cos ^2 \theta$ and $\Delta=r^2-2Mr+a^2+Q^2$. Here $M$,
$Q$ and $a$ are the total mass, total charge and specific angular momentum
respectively characterizing the spacetime.  The (outer) event horizon is
located at $r_+=M+\sqrt{M^2-a^2-Q^2}$.

Let us introduce the Carter's observer family \cite{carter}, whose four-velocity is given by
\begin{equation}
\label{Ucarter}
U_{\rm car}= \frac{r^2+a^2}{\sqrt{\Delta \, \Sigma}} \left[\partial_t + \frac{a}{r^2+a^2}\partial_\phi\right]\ .
\end{equation}
An observer adapted frame to $U_{\rm car}$ is then easily constructed with the triad
\begin{eqnarray}
\label{carterframe}
e_{\hat r}&=&\frac{1}{\sqrt{g_{rr}}}\partial_r\ , \qquad
e_{\hat \theta}=\frac{1}{\sqrt{g_{\theta\theta}}}\partial_\theta\ , \nonumber\\
{\bar U}_{\rm car}&=&\frac{a\sin \theta}{\sqrt{\Sigma}} \left[\partial_t + \frac{1}{a\sin^2\theta }\partial_\phi\right]\ .
\end{eqnarray}
The Carter observers measure parallel electric and magnetic fields $E$ and $B$ \cite{damruff}, with components
\begin{equation}
E(U_{\rm car})^\alpha=F^\alpha{}_\beta U_{\rm car}^\beta\ , \qquad
B(U_{\rm car})^\alpha={}^*F^\alpha{}_\beta U_{\rm car}^\beta\ ,
\end{equation}
where ${}^*F$ is the dual of the electromagnetic field (\ref{Fkn}).
Both $E$ and $B$ are directed along $e_{\hat r}$ and assuming as usual $Q>0$, the strength of electric and magnetic fields are given by
\begin{eqnarray}
\label{EBsols} |{\bf E}|&=&|E^{\hat r}|=\frac{Q}{\Sigma^2}(r^2-a^2\cos^2\theta)\
, \nonumber\\ 
|{\bf B}|&=&|B^{\hat r}|=\left|2\frac{Q}{\Sigma^2}ar\cos\theta\right|\ .
\end{eqnarray}
It is worth noting that the Carter orthonormal frame is the unique frame in which the flat spacetime Schwinger discussion can be locally applied.
This is due both to the meaning of the Carter orthonormal frame and its relation to the geometry of the Weyl curvature tensor and the spacetime itself, as well as to the fact that the invariantly described Schwinger process demands this unique frame for its application.
An alternative but equivalent derivation of this result is presented in
Appendix A, where the electric and magnetic field strengths are obtained in
terms of the electromagnetic invariants by using the Newman-Penrose formalism,
hence showing more clearly the invariant character of the dyadotorus.

The Schwinger formula generalized to include both electric and magnetic fields, i.e.
\begin{eqnarray}
\label{schwingerKN}
2{\rm Im}\mathcal L&=&\frac1{4\pi}\left(\frac{|{\bf E}|e}{\pi\hbar}\right)^2\sum_{n=1}^\infty\frac1{n^2}\left(n\pi\frac{|{\bf B}|}{|{\bf E}|}\right)\nonumber\\
&&\times\coth\left(n\pi\frac{|{\bf B}|}{|{\bf E}|}\right)e^{-n\pi E_c/|{\bf E}|}\ ,
\end{eqnarray}
has been used by Damour and Ruffini \cite{damruff} for the case of a Kerr-Newman geometry.

\begin{figure*}
\typeout{*** EPS figure 2}
\begin{center}
$\begin{array}{cc}
\includegraphics[scale=0.45]{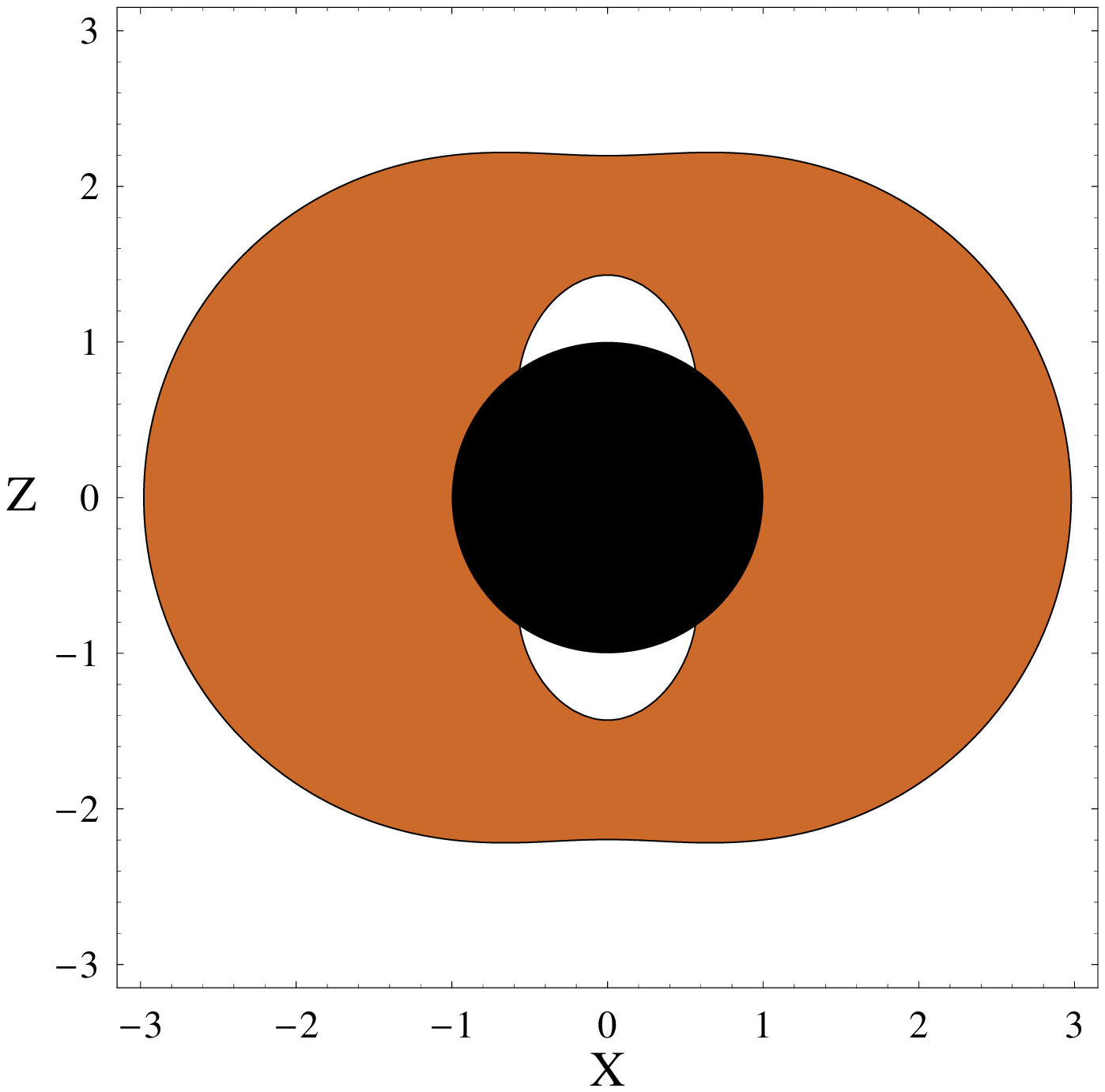}&\qquad
\includegraphics[scale=0.45]{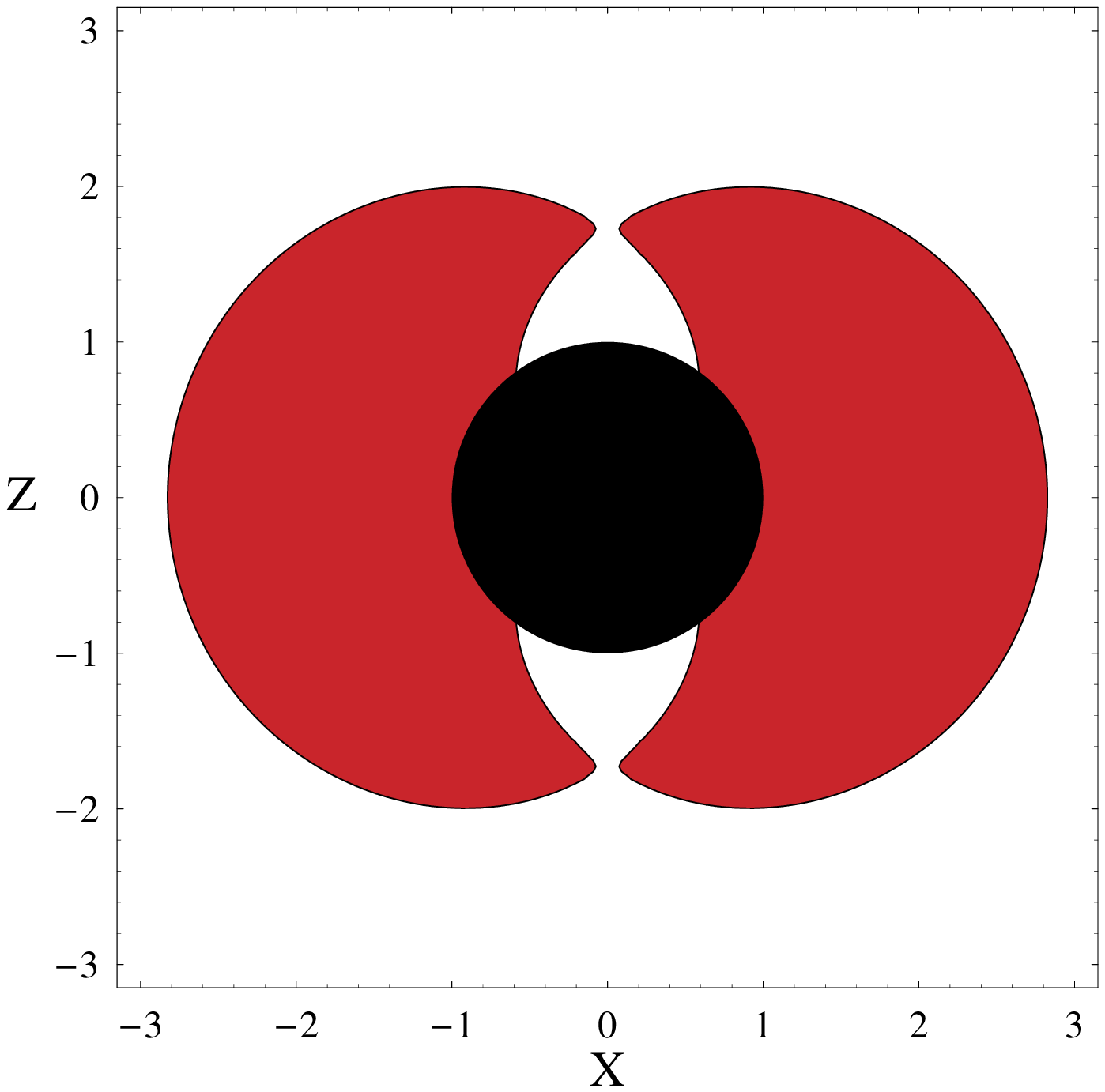}\\[0.4cm]
\mbox{(a)} & \mbox{(b)}\\[0.6cm]
\includegraphics[scale=0.45]{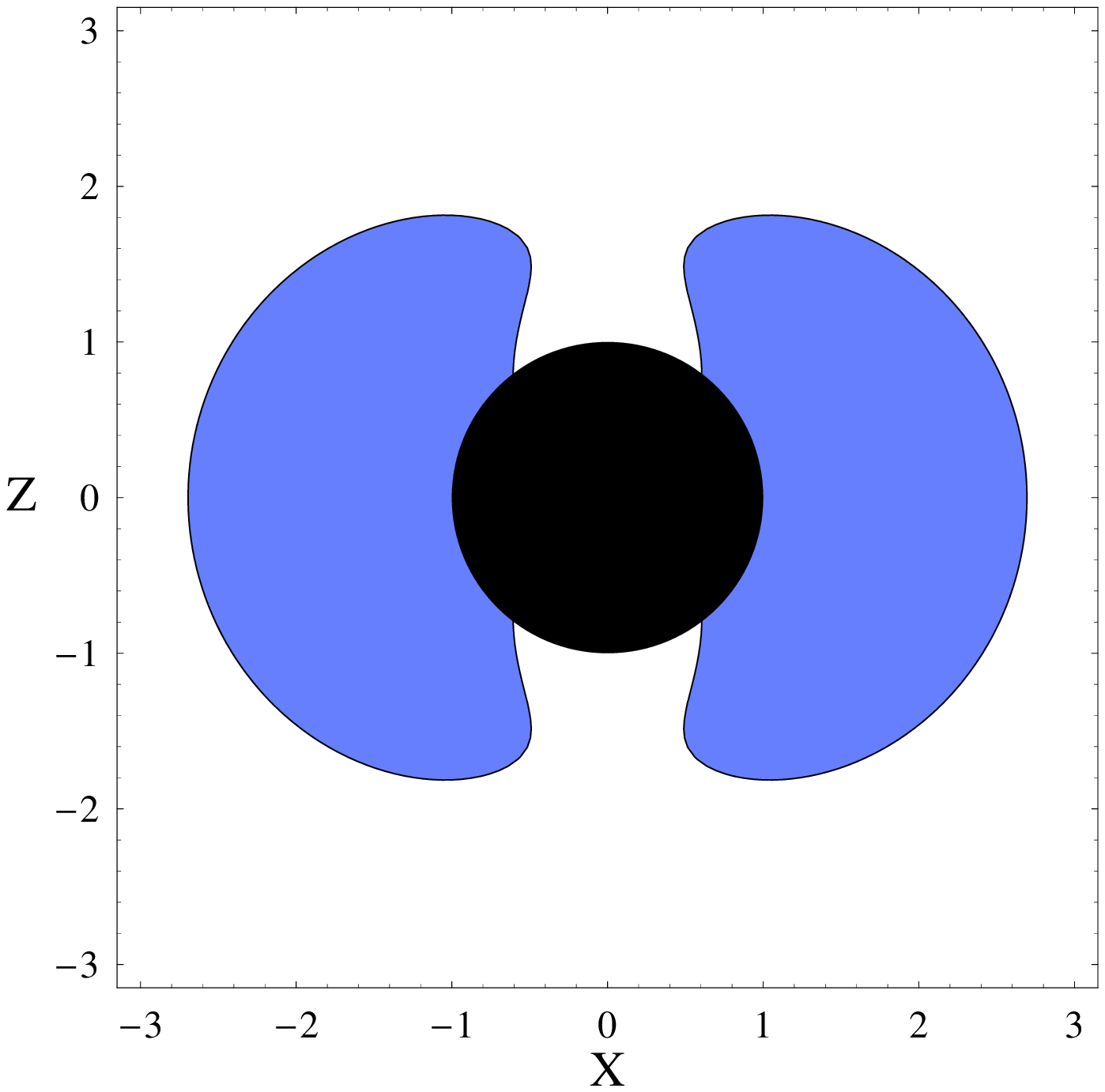}&\qquad
\includegraphics[scale=0.45]{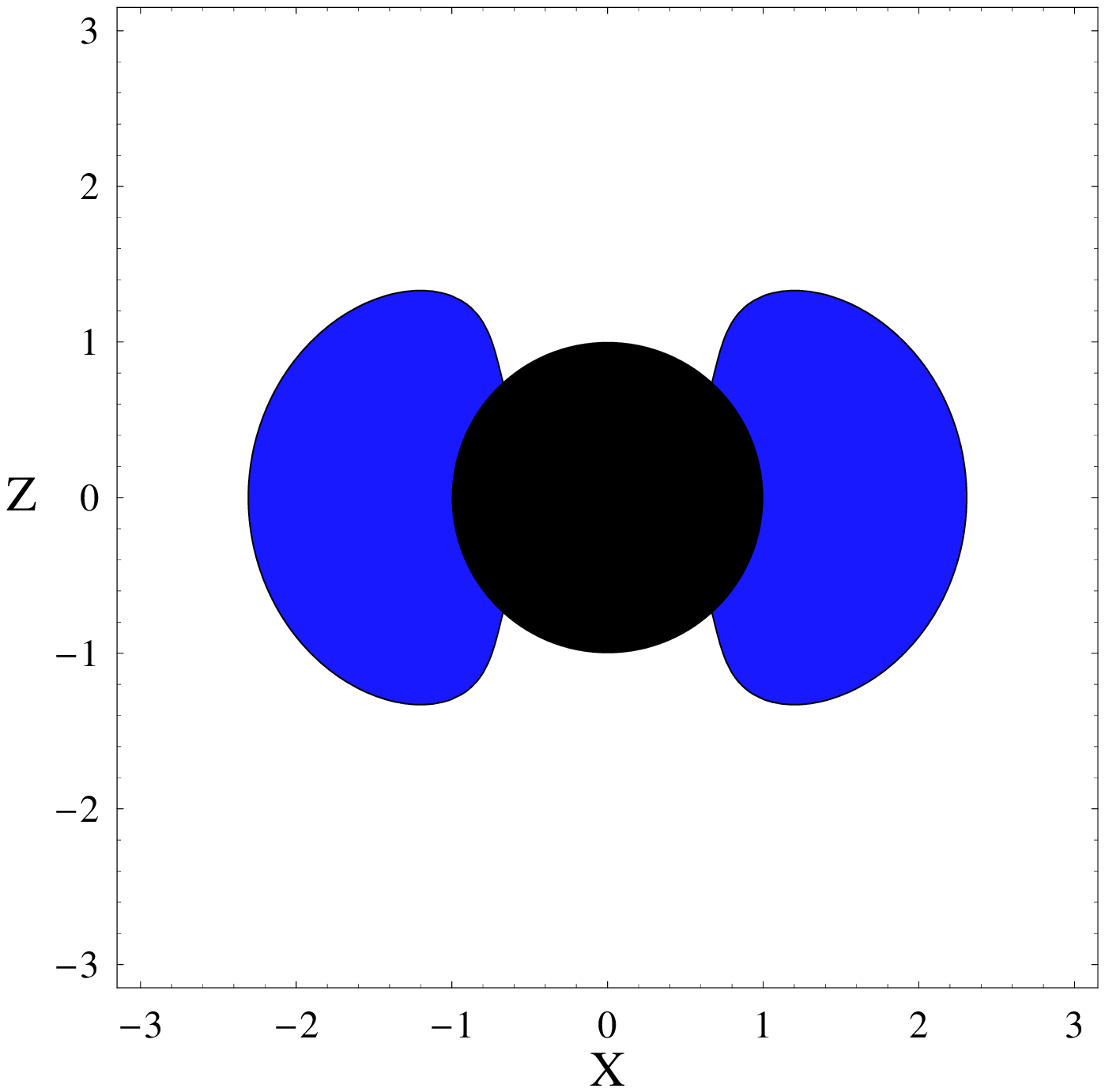}\\[0.4cm]
\mbox{(c)} & \mbox{(d)}\\
\end{array}$
\end{center}
\caption{The projection of the dyadotorus on the $X-Z$ plane ($X=r\sin \theta$, $Z=r\cos \theta$ are Cartesian-like coordinates built up simply using the Boyer-Lindquist radial and angular coordinates) is shown for an extreme
Kerr-Newman black hole with $\mu=10$, $\lambda=1.49\times10^{-4}$ and different values of the parameter $k$: (a) $k=0.9$ (orange), (b) $k=1.0$ (red), (c) $k=1.1$ (light blue), (d) $k=1.5$ (blue).
The boundary of the dyadoregion becomes a torus-like surface for $k\approx0.998$, according to Eq. (\ref{condxi}).
The black disk represents the black hole horizon.
}
\label{fig:2}
\end{figure*}

We are interested in the region exterior to the outer horizon $r\geq r_+$.
Solving Eq. (\ref{dyadoredef}) for $r$ and introducing the dimensionless
quantities $\lambda=Q/M$, $\alpha=a/M$, $\mu=M/M_\odot$ and $\epsilon=k E_c
M_\odot\approx1.873 k \times10^{-6}$ (with $M_\odot\approx1.477\times10^{5}$
cm) we get
\begin{equation}
\label{dyadosurf}
\left(\frac{r^d_\pm}{M}\right)^2=\frac12\frac{\lambda}{\mu\epsilon}
-\alpha^2\cos^2\theta\pm\left[\frac14\frac{\lambda^2}{\mu^2\epsilon^2}
-2\frac{\lambda}{\mu\epsilon}\alpha^2\cos^2\theta\right]^{1/2}\,
\end{equation}
where the $\pm$ signs correspond to the two different parts of the surface.
They join at the particular values $\theta^*$ and $\pi-\theta^*$ of the polar
angle given by the condition of vanishing argument of the square root in Eq.
(\ref{dyadosurf})
\begin{equation}
\theta^*=\arccos\left(\frac{1}{2\sqrt{2}\alpha}\sqrt{\frac{\lambda}{\mu\epsilon}}\right)\ .
\end{equation}
The requirement that $\cos\theta^*\leq1$ can be solved for instance for the
constant parameter $k$, giving the range of allowed values for which the
dyadotorus appears indeed as a torus-like surface (see Figs. \ref{fig:2} (b),
(c) and (d))
\begin{equation}
\label{condxi}
k \geq \frac{\lambda}{8 E_c M_\odot \mu \alpha^2}\approx 6.6\times 10^4
\frac{\lambda}{\mu \alpha^2}\ ;
\end{equation}
for lower values of $k$ the dyadotorus consists instead of two disjoint parts,
one of them (corresponding to the branch $r^d_+$) always external to the other
(corresponding to the branch $r^d_-$), and has rather the shape of an ellipsoid
(see Fig. \ref{fig:2} (a)). Therefore, the use of the term dyadoregion should
be more appropriate in this case.

\begin{figure}
\typeout{*** EPS figure 1}
\begin{center}
\includegraphics[scale=0.45]{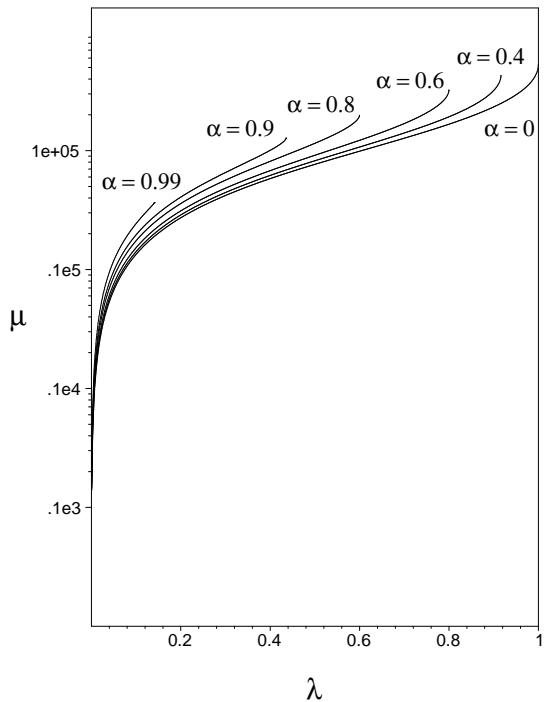}
\caption{The space of parameters $(\lambda,\mu)$ is shown for different values of the rotation parameter $\alpha=a/M=[0,0.4,0.6,0.8,0.9,0.99]$ and fixed value of the polar angle $\theta=\pi/3$.
The region below each curve represents the allowed region for the existence of the dyadoregion with fixed $\alpha$.
The configurations above each line correspond to unphysical solutions where $r^d_\pm<r_+$ for the selected set of parameters. The value of the parameter $k$ has been set equal to one.
}
\label{fig:1}
\end{center}
\end{figure}

In terms of the dimensionless quantities $\lambda$ and $\alpha$ the horizon radius is then given by
\begin{equation}
\frac{r_+}{M}=1+\sqrt{1-\lambda^2-\alpha^2}\ .
\end{equation}
The condition for the existence of the dyadotorus is given by $r^d_\pm\geq
r_+$. The allowed region for the pairs $(\lambda,\mu)$ (with fixed values of the
rotation parameter $\alpha$ and the polar angle $\theta$) satisfying this
condition is shown in Fig. \ref{fig:1}.


Figure \ref{fig:2} shows the shape of the projection of the dyadotorus on a plane containing the rotation axis for an extreme Kerr-Newman black hole with fixed $\mu$ and $\lambda$ and different values of the parameter $k$ using Cartesian-like coordinates $X=r\sin \theta$, $Z=r\cos \theta$, built up simply by taking the Boyer-Lindquist coordinates $r$ and $\theta$ as polar coordinates in flat space.


\begin{figure}
\typeout{*** EPS figure 2 bis}
\begin{center}
$\begin{array}{c}
\includegraphics[scale=1]{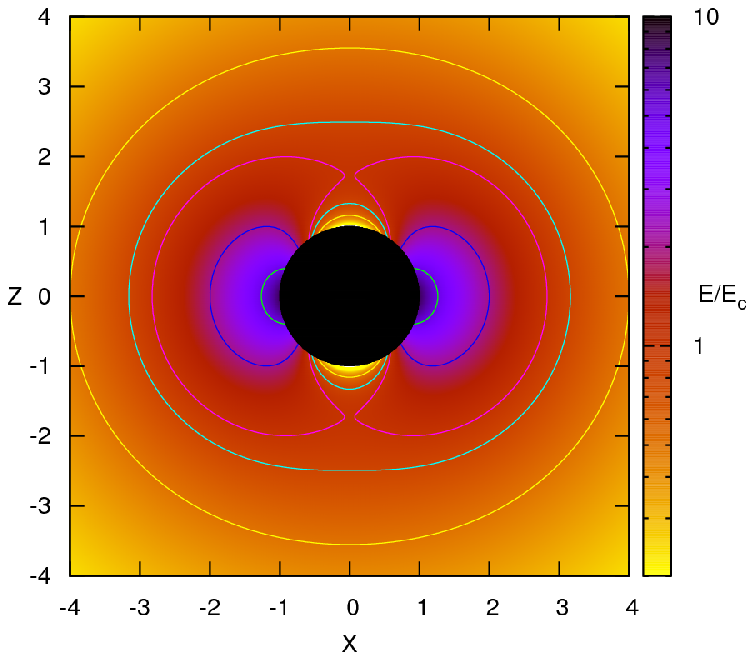} \\
[0.4cm] \mbox{(a)} \\[0.6cm]
\includegraphics[scale=1]{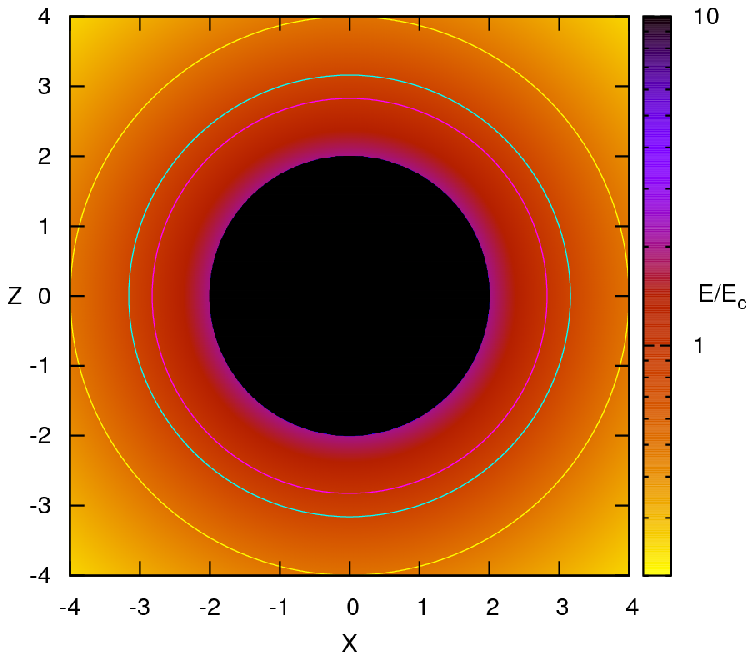}\\[0.4cm]
\mbox{(b)}\\
\end{array}$
\end{center}
\caption{The projections of the dyadotorus on the $X-Z$ plane corresponding to
different values of the ratio $|{\bf E}|/E_c\equiv k$ are shown in Fig. (a) for
$\mu=10$ and $\lambda=1.49\times 10^{-4}$.  The corresponding plot
for the dyadosphere with the same mass energy and charge to mass ratio is shown in Fig.
(b) for comparison.
}
\label{fig:2bis}
\end{figure}

A \lq\lq dynamical'' view of topology change in the shape of the dyadoregion is shown in Fig. \ref{fig:2bis}, where the case of a Reissner-Nordstr\"{o}m black hole with the same total mass and charge is also shown for comparison.
We point out some interesting qualitative differences between dyadotorus and dyadosphere which can be seen clearly from these plots.  In particular, the dyadotorus appear to lead to larger values of the electric field than the corresponding
dyadosphere close to the horizon.
A key point here is the size of the horizon, which in the limit of small charge to mass ratio $\lambda\ll1$ for an extreme Kerr-Newman black hole goes to $r_+ \sim M$, while in the case of a Reissner-Nordstr\"{o}m black hole goes to $r_+\sim 2M$.
This fact is crucial because it leads to the presence of stronger electric fields for the Kerr-Newman black hole in contrast with the Reissner-Nordstr\"om one.
We can compare for instance the maximum electric field $E_{\rm max}=Q/r^2_+$ of an extreme Kerr-Newman black hole and
of a Reissner-Nordstr\"{o}m black hole, which is obtained for $r=r_+$, $\theta=\pi/2$ in the former case and $r=r_+$ in the latter case, in the limit of small charge to mass ratio
\begin{equation}
E^{\rm KN}_{\rm max}= \frac{Q}{M^2} = 4 E^{\rm RN}_{\rm max}\ .
\end{equation}
We will turn to the energetics of the dyadoregion in Section V.


\begin{figure*}
\typeout{*** EPS figure 3}
\begin{center}
$\begin{array}{cc}
\includegraphics[scale=0.45]{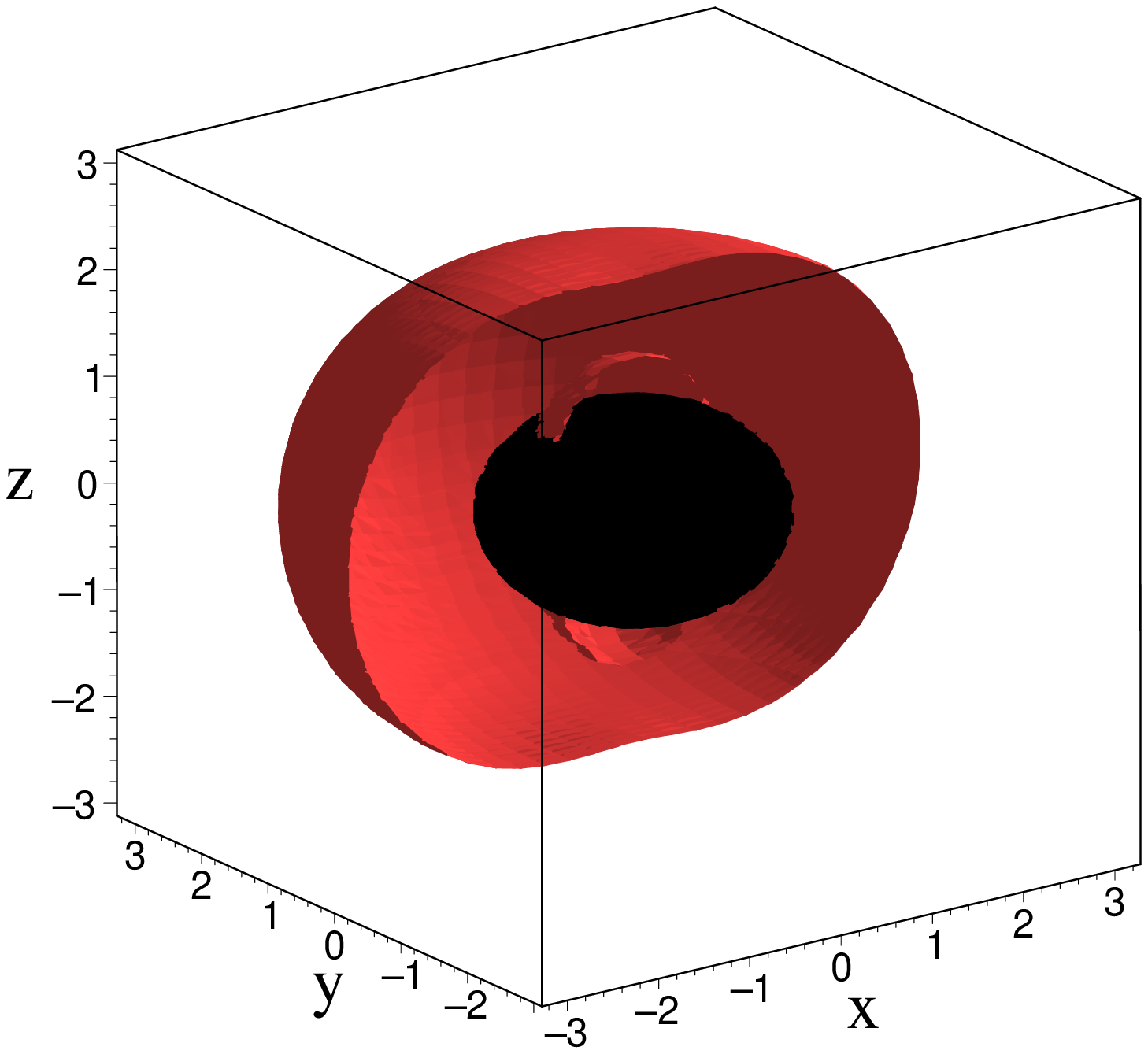}&\qquad
\includegraphics[scale=0.45]{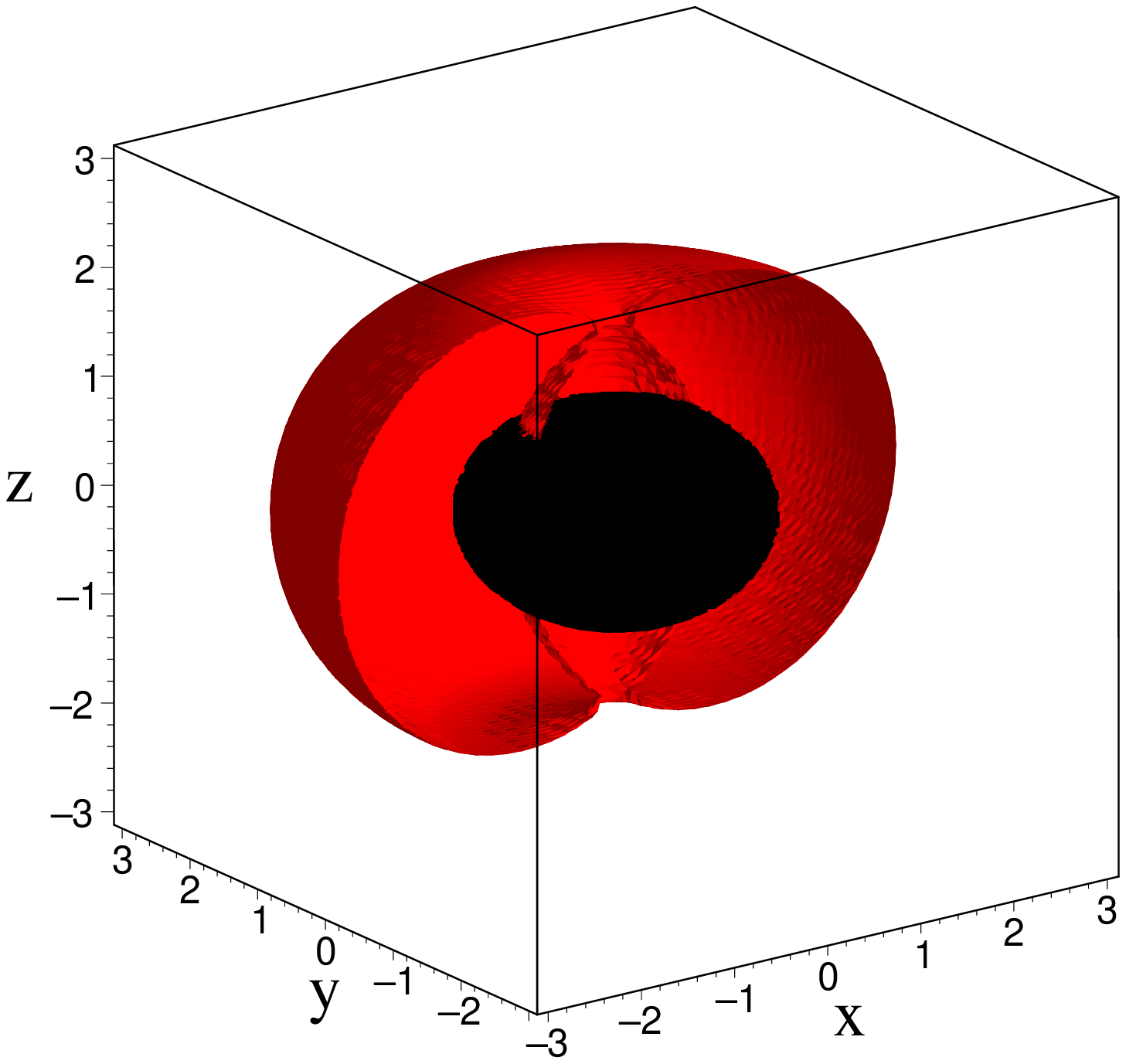}\\[0.4cm]
\mbox{(a)} & \mbox{(b)}\\[0.6cm]
\includegraphics[scale=0.45]{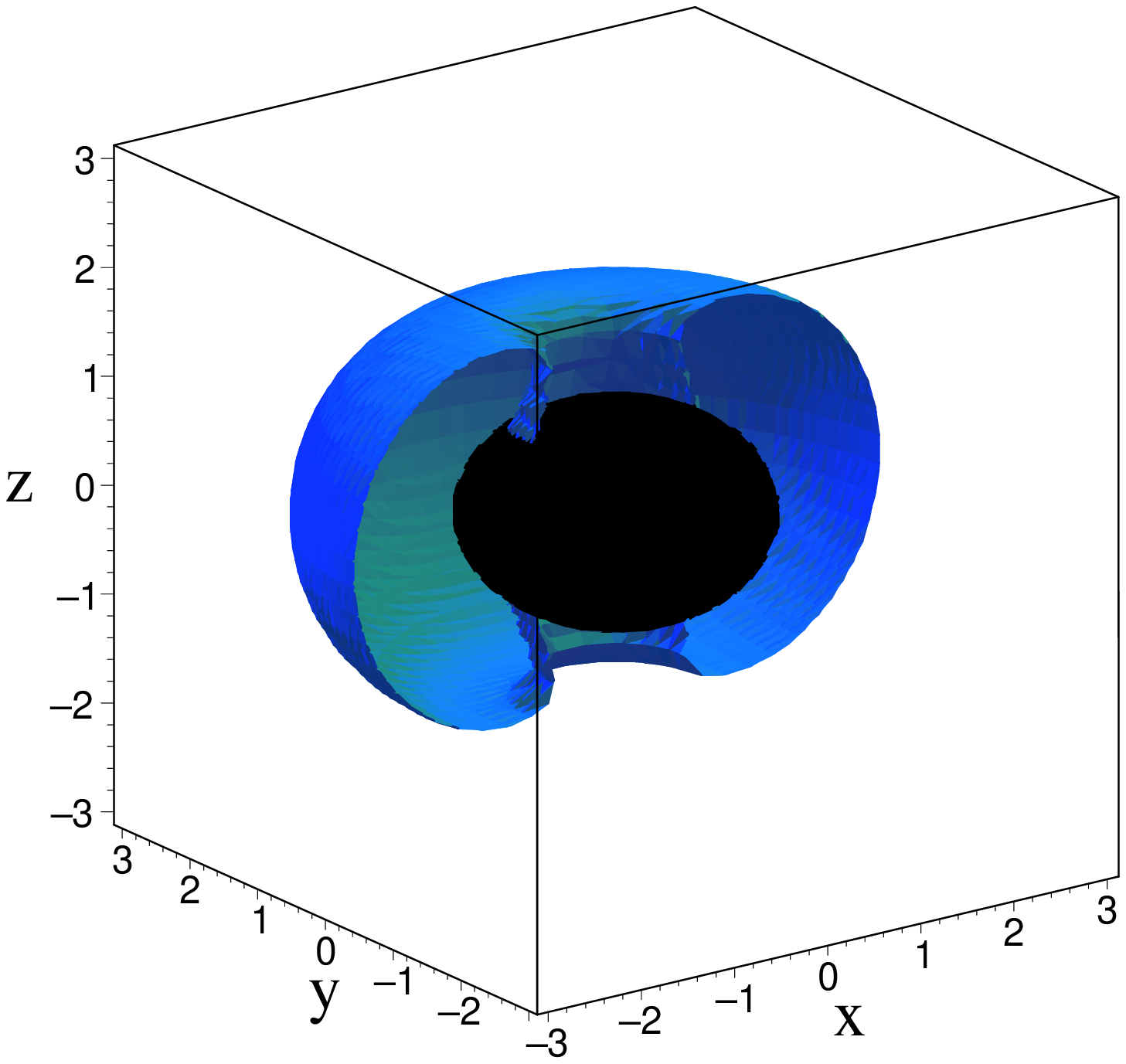}&\qquad
\includegraphics[scale=0.45]{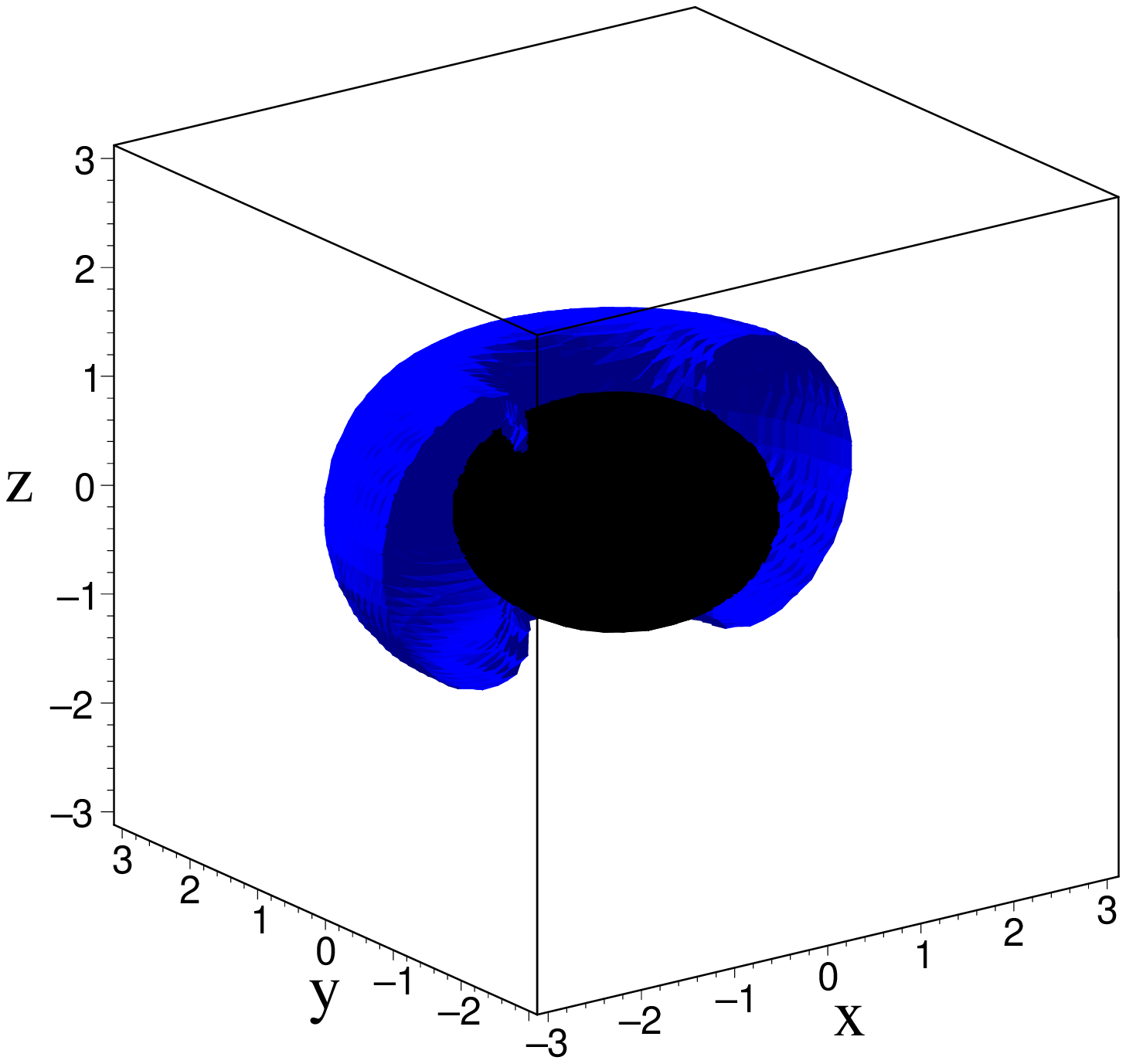}\\[0.4cm]
\mbox{(c)} & \mbox{(d)}\\
\end{array}$
\end{center}
\caption{Three-dimensional images of the dyadotorus are shown using
Kerr-Schild coordinates.  The parameter choice is the same as in Fig.
\ref{fig:2}. The surfaces have been cut in half for a better view of the
interior. The horizon instead has been shown entirely (black region).
}
\label{fig:3}
\end{figure*}

Three-dimensional images of the dyadotorus can be generated also in terms of
Kerr-Schild coordinates $(\tilde t,x,y,z)$, which are related to the standard
Boyer-Lindquist ones $(t,r,\theta,\phi)$ by the equations (see e.g. \cite{MTW})
\begin{eqnarray}
\label{kerrschild}
d\tilde t&=&dt-\frac{2Mr-Q^2}{\Delta}dr\ , \nonumber\\
d\psi&=&d\phi-\frac{a}{r^2+a^2}\frac{2Mr-Q^2}{\Delta}dr\ , \nonumber\\
x&=&\sqrt{r^2+a^2}\sin\theta\cos\psi\ , \nonumber\\
y&=&\sqrt{r^2+a^2}\sin\theta\sin\psi\ , \nonumber\\
z&=&r\cos\theta\ .
\end{eqnarray}
Note that the spatial coordinates $(x,y,z)$ satisfy the relation
\begin{equation}
\frac{x^2+y^2}{r^2+a^2}+\frac{z^2}{r^2}=1\ ,
\end{equation}
and the auxiliary angular coordinate $\psi$ is a function of $r$, as from the second relation of Eq. (\ref{kerrschild})
\begin{equation}
\psi=\phi-\int^r\frac{a}{r^2+a^2}\frac{2Mr-Q^2}{\Delta}dr\ .
\end{equation}
The shape of the dyadotorus using Kerr-Schild coordinates is shown in Fig.
\ref{fig:3} for the same choice of parameters as in Fig. \ref{fig:2}.


\section{Embedding diagram}

The plots of Figs.~\ref{fig:2} and \ref{fig:3} actually shows a distorted view of the shape of the dyadotorus; we should rather look at the corresponding embedding diagram, which gives the correct geometry allowing to visualize the spacetime curvature.
Because of our familiar three-dimensional intuition, the most useful and easily understood embedding diagrams are those which take a Riemannian two-surface from the original geometry, then reconstructing it as a distorted surface in a three-dimensional Euclidean space.

The dyadotorus implicitly defined by Eq. (\ref{dyadosurf}) can be visualized as a 2-dimensional surface of revolution around the rotation axis embedded in the usual Euclidean 3-space by suppressing the temporal and azimuthal dependence.
Using Boyer-Lindquist coordinates, form Eq. (\ref{KNmetric}) we get the following induced metric of the constant time slice ($dt=0$) of the world sheet $r=r^d_\pm$ given by Eq. (\ref{dyadosurf})
\begin{eqnarray}
\label{embmetric}
^{(2)}ds^2&=&h_{\eta\eta}d\eta^2+g_{\phi\phi}d\phi^2\ , \nonumber\\
h_{\eta\eta}&=&g_{rr}\left(\frac{dr^d_\pm}{d\eta}\right)^2+\frac{g_{\theta\theta}}{1-\eta^2}\ ,
\end{eqnarray}
where $\eta\equiv\cos \theta$ and all the metric coefficients are evaluated at $r=r^d_\pm$, which is indeed a function of the polar angle $\theta$ (so that $dr$ has been related to $d\eta$).

Following a standard procedure \cite{smarr,sharp}, consider the flat-space line element written in spherical-like coordinates
\begin{equation}
\label{3euclmetric}
^{(3)}ds^2=dX^2+dY^2\pm dZ^2\ ,
\end{equation}
where the plus sign refers to the Euclidean case and the minus sign to the Minkowskian case.
For the embedding surface in the parametric form
\begin{equation}
\label{embcoords}
X=F(\eta)\cos\phi\ , \qquad Y=F(\eta)\sin\phi\ , \qquad Z=G(\eta)\ ,
\end{equation}
the corresponding line element becomes
\begin{equation}
\label{2euclmetric}
^{(2)}ds^2=\left[\left(\frac{dF}{d\eta}\right)^2\pm\left(\frac{dG}{d\eta}\right)^2\right]d\eta^2+F^2d\phi^2\ .
\end{equation}
Comparison with (\ref{embmetric}) implies
\begin{equation}
\left(\frac{dF}{d\eta}\right)^2\pm\left(\frac{dG}{d\eta}\right)^2=h_{\eta\eta}\ , \qquad F=\sqrt{g_{\phi\phi}}\ .
\end{equation}
The relation $F=F(\eta)$ is already given by the second equation and one can then numerically integrate the first equation to get the function $G(\eta)$
\begin{equation}
\label{Zdicsol}
G_\pm(\eta)=\int_{\eta_0}^\eta\sqrt{\pm\left(h_{\eta\eta}-\left[\frac{d}{d\eta}(\sqrt{g_{\phi\phi}})\right]^2\right)}\,d\eta\ ,
\end{equation}
with the initial condition $G(\eta_0)=0$.
Note that the dyadotorus is embeddable entirely in the Euclidean 3-space, whereas the embedding of the outer horizon may become Minkowskian depending on the values of the charge and rotation parameters of the black hole \cite{smarr}.
In the latter case the embedding cross section has a horizontal tangent line when the signature switch of sign in Eq. (\ref{Zdicsol}) takes place at a certain value of $\eta$ given by $\eta=\eta_{\rm(ss)}$, where $dG/d\eta=0$.
The integration must be performed with the plus sign (into the
Euclidean part of the embedding) or with the minus sign (into the Minkowskian part of the
embedding) starting from such a signature-switch point with the initial condition $G(\eta_{\rm(ss)})=0$.

\begin{figure*}
\typeout{*** EPS figure 4}
\begin{center}
$\begin{array}{cc}
\includegraphics[scale=0.45]{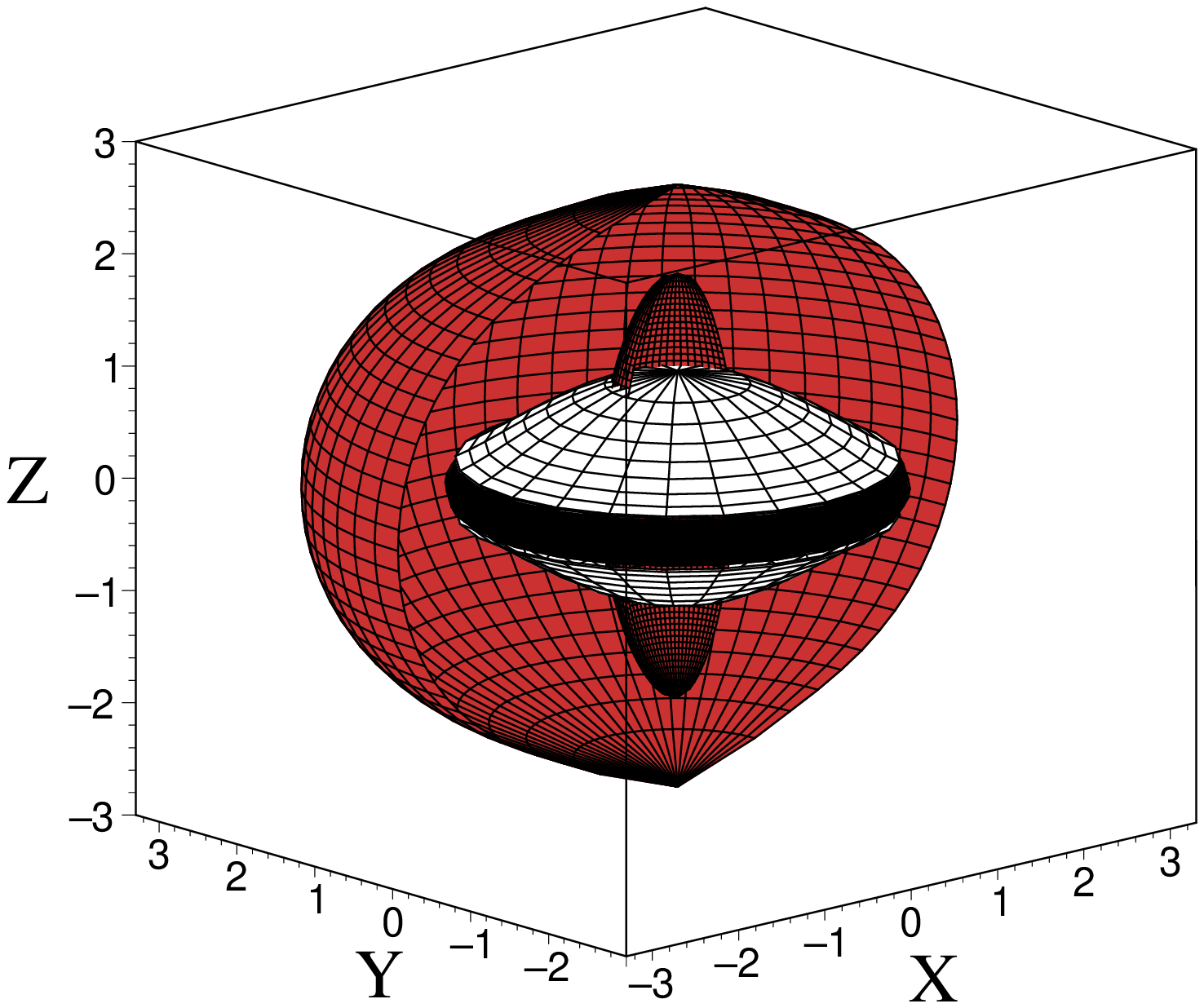}&\qquad
\includegraphics[scale=0.45]{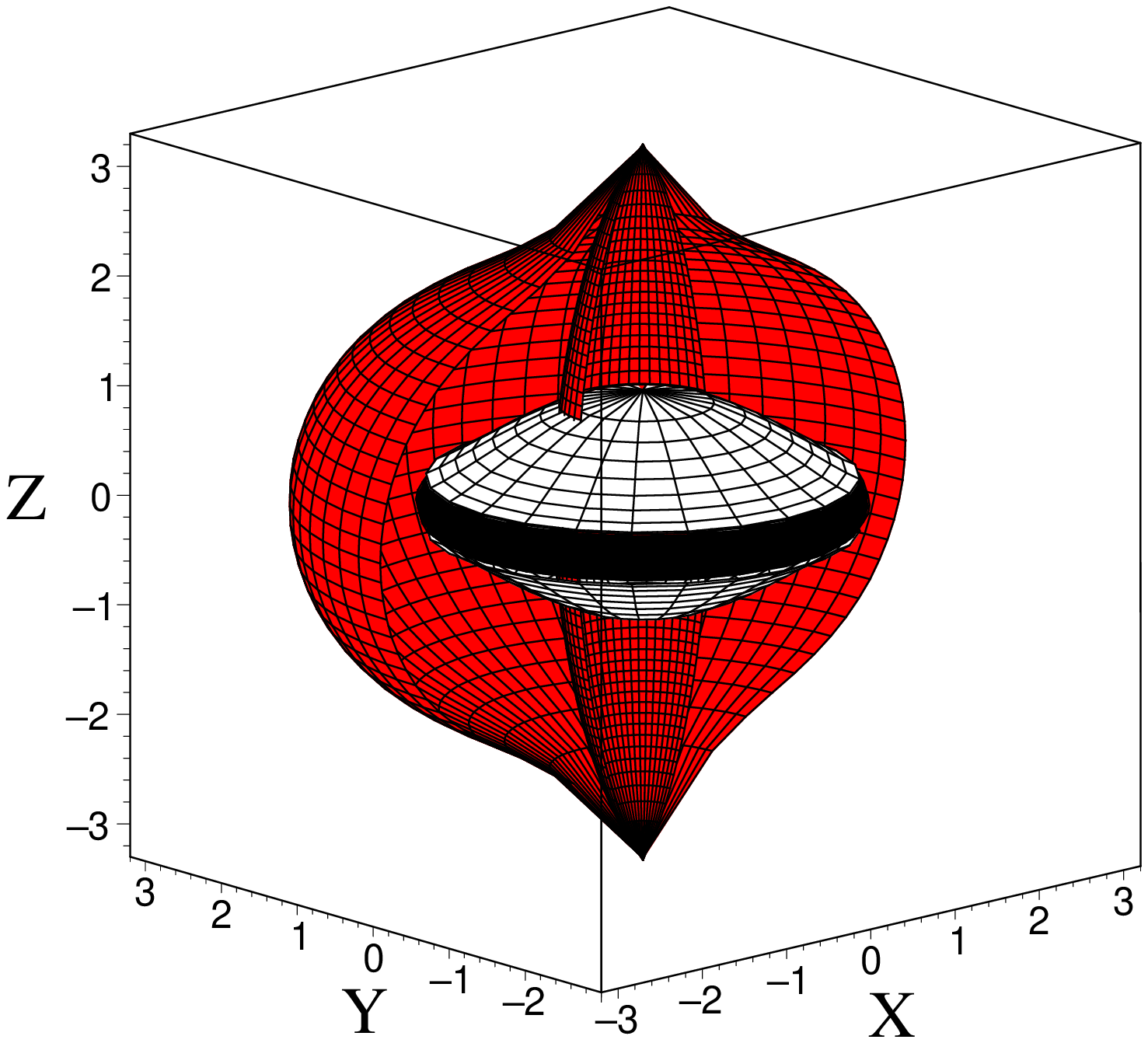}\\[0.4cm]
\mbox{(a)} & \mbox{(b)}\\[0.6cm]
\includegraphics[scale=0.45]{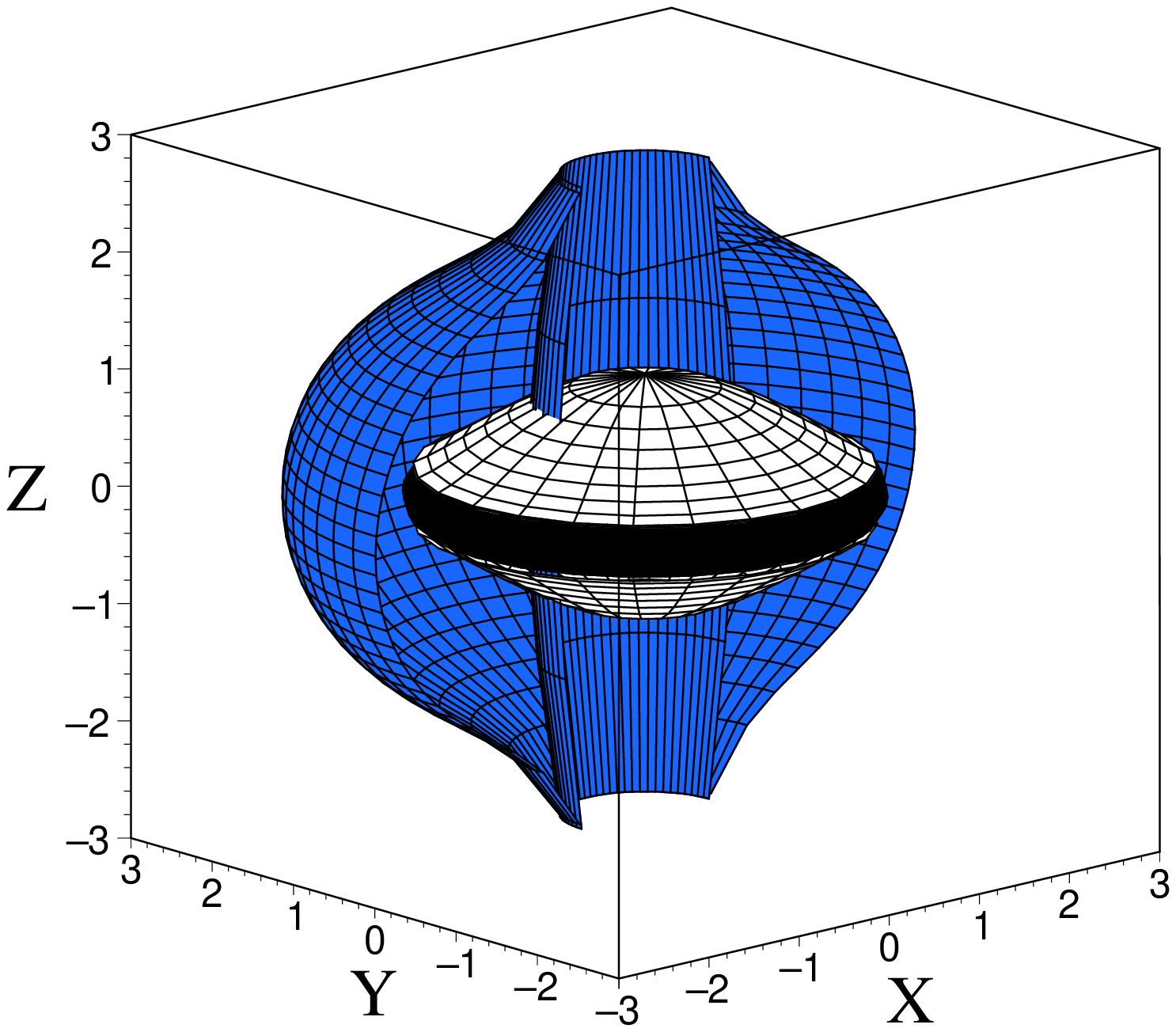}&\qquad
\includegraphics[scale=0.45]{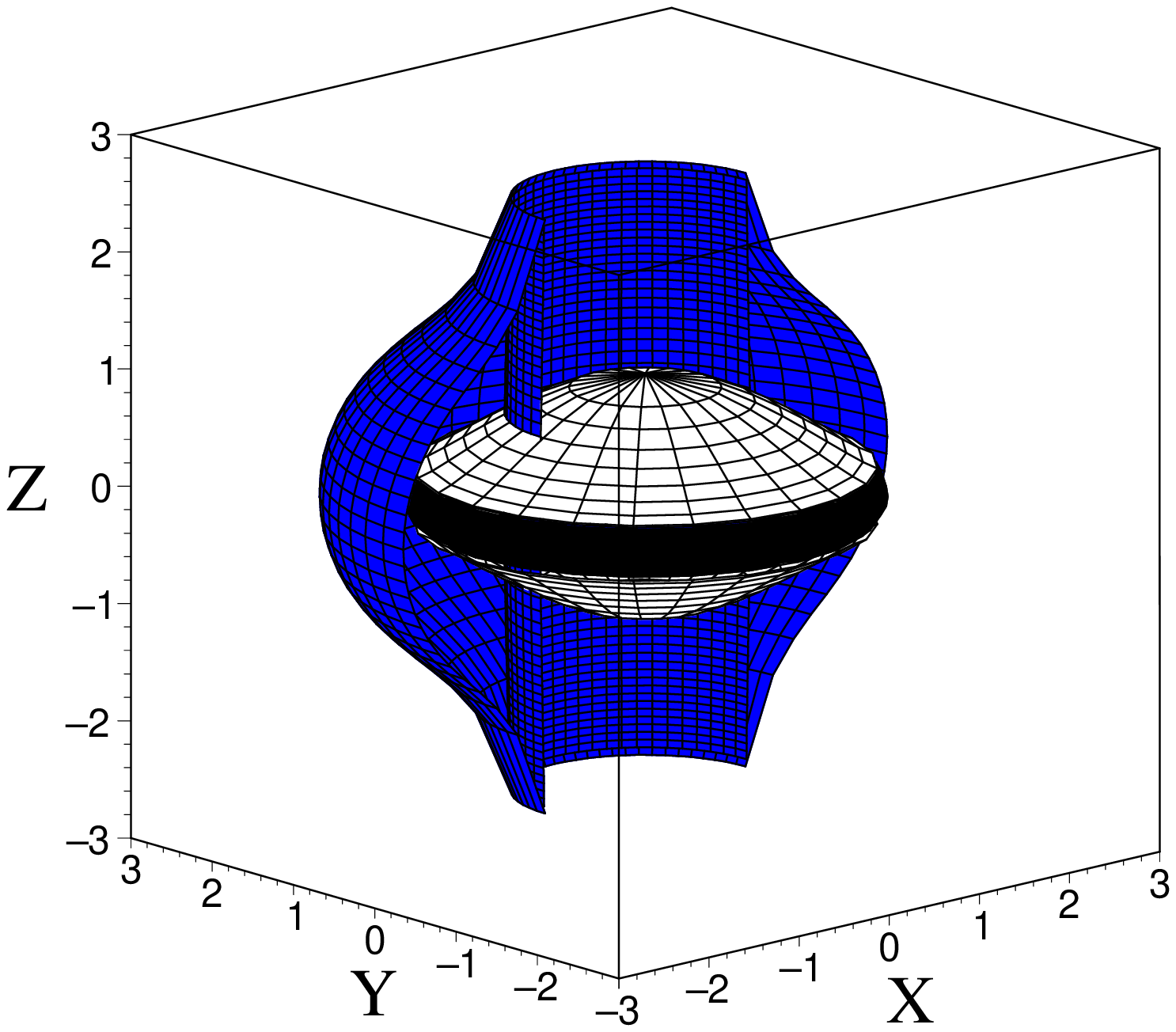}\\[0.4cm]
\mbox{(c)} & \mbox{(d)}\\
\end{array}$
\end{center}
\caption{The dyadotorus is shown on an embedding diagram. The choice of the
parameters is the same as in Fig.~\ref{fig:2} as concerns
Figs. (a), (c) and (d). In the case of Fig. (b) the value of the parameter $k$
has been changed to the critical value $k\approx0.998$ in order to satisfy the condition
(\ref{condxi}) with the equality sign, so representing the limiting case in
which the dyadotorus still appears as a torus-like surface (as in Figs. (c)
and (d)) for the chosen values of parameters $\mu$ and $\alpha$. The surfaces
have been cut in half for a better view of the interior, where the embedding of
the horizon is also shown (the black shaded region is Euclidean, whereas the
white regions are Minkowskian). Note that in this case the coordinates
$(X,Y,Z)$ are given by Eq. (\ref{embcoords}).
}
\label{fig:4}
\end{figure*}

Fig. \ref{fig:4} shows the embedding diagram of the dyadotorus for the same choice of parameters as in Figs.~\ref{fig:2} and \ref{fig:3} as concerns Figs. (a), (c) and (d). Fig. (b) corresponds instead to a slightly different choice of the charge parameter, satisfying Eq. (\ref{condxi}) with the equality sign (implying $\theta^*=0$), i.e. to the limiting value of $k$ such that the dyadotorus still appears as a torus-like surface, the two branches $r^d_\pm$ still joining (at $\theta=0,\pi$).
Despite the appearance the cusps on the axis do not correspond to conical singularities at the axis ($\theta=0,\pi$), as it occurs in contrast in the case of the ergosphere \cite{sharp,lake}.
In fact, expanding the induced metric (\ref{embmetric}) about $\theta=0$ (or equivalently $\theta=\pi$) to the second order we get the approximate metric (up to an ignorable constant factor)
\begin{equation}
^{(2)}ds^2\simeq d\theta^2+\theta^2d\phi^2\
\end{equation}
with $\phi\in[0,2\pi]$, which is the intrinsic metric of a right cone with no deficit angle.

\begin{figure*}
\typeout{*** EPS figure 5}
\begin{center}
$\begin{array}{cc}
\includegraphics[scale=0.45]{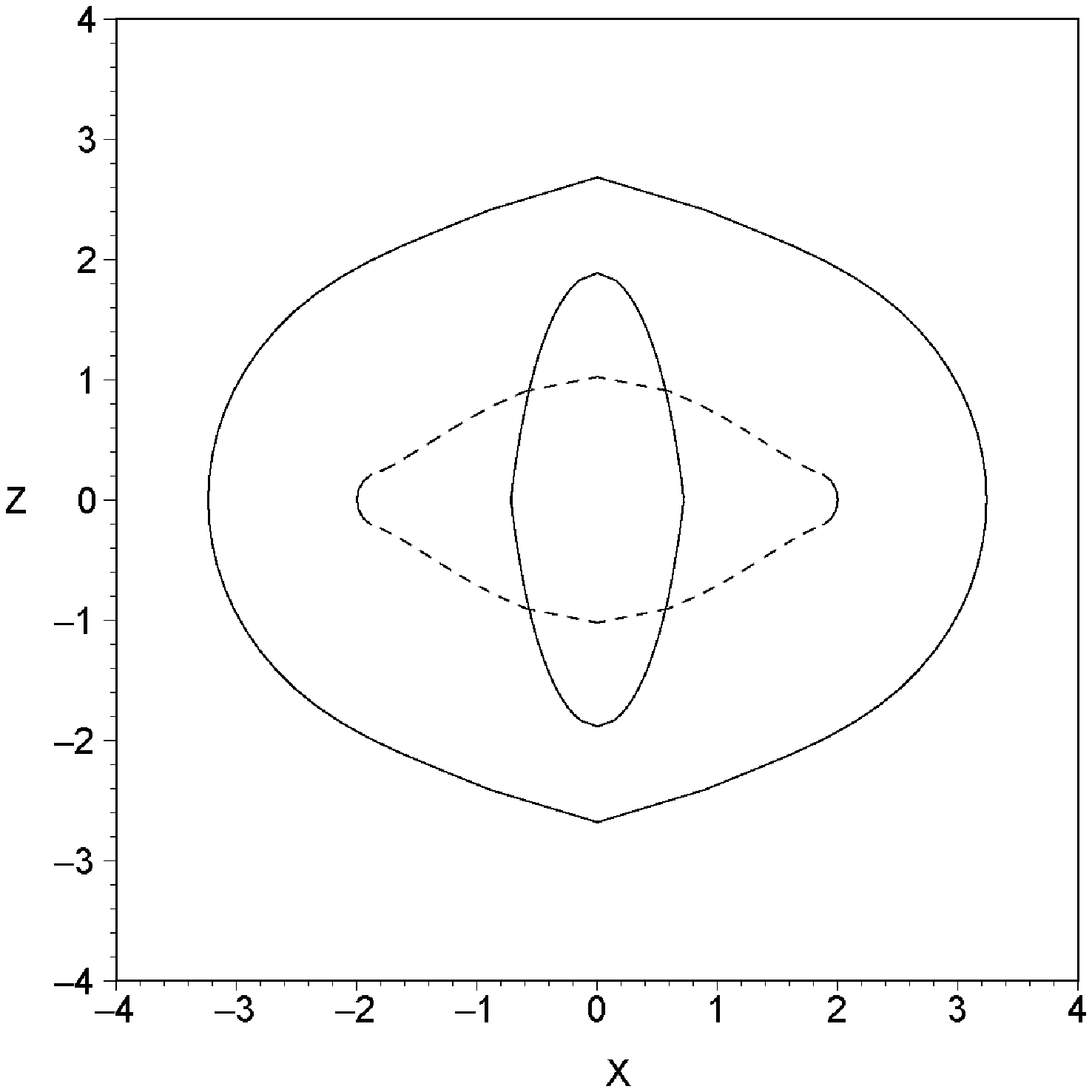}&
\includegraphics[scale=0.45]{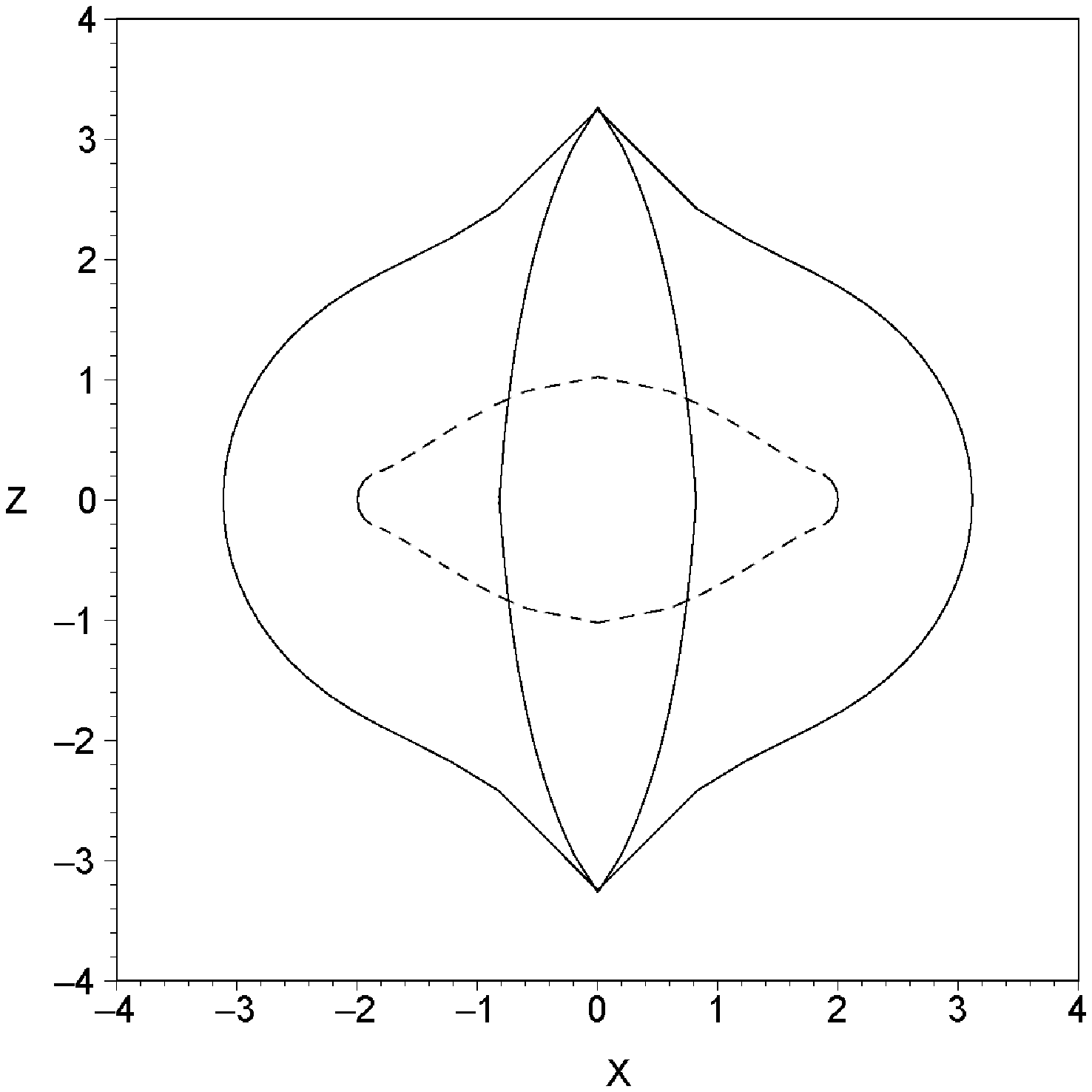}\\[.4cm]
\mbox{(a)} & \mbox{(b)}\\[.6cm]
\includegraphics[scale=0.45]{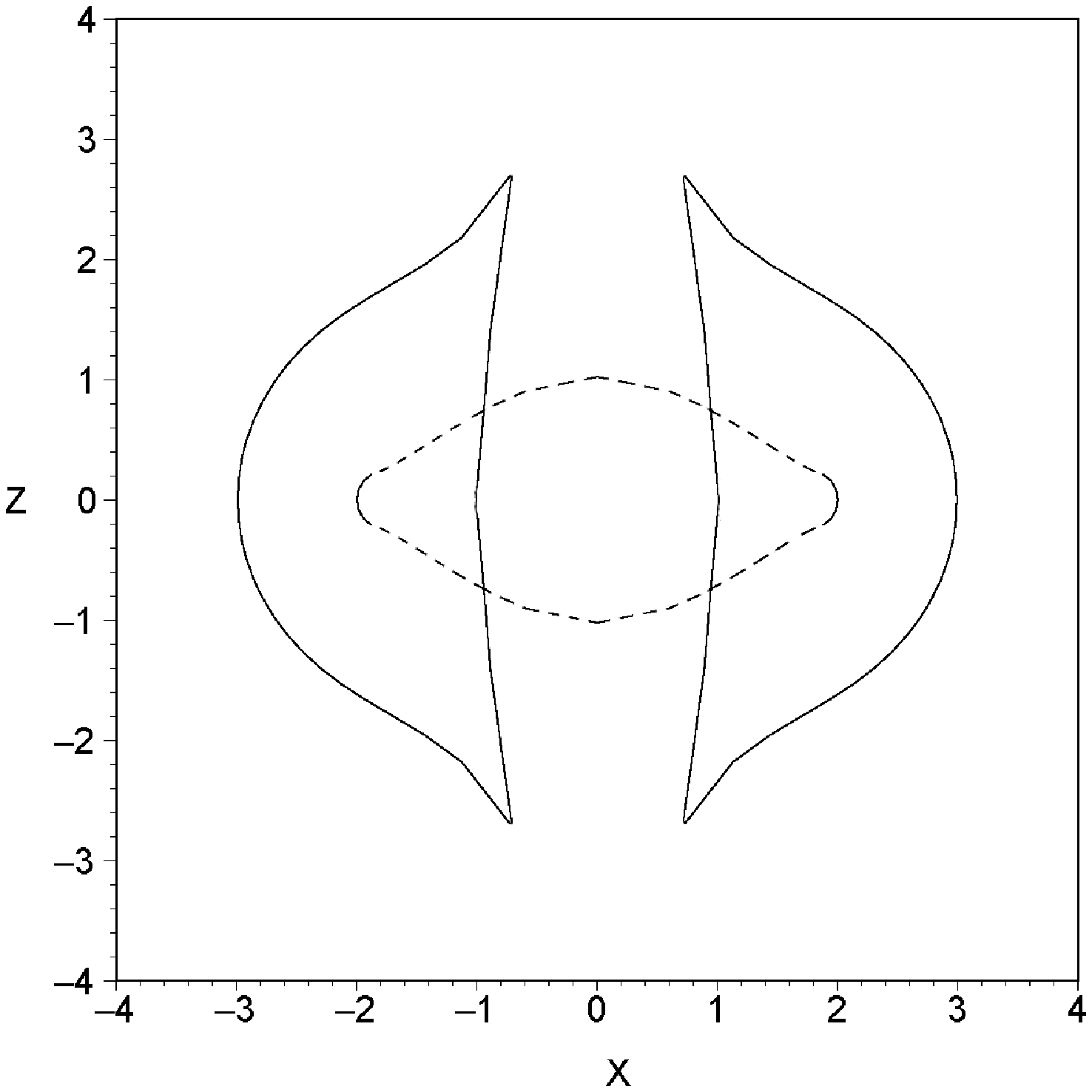}&
\includegraphics[scale=0.45]{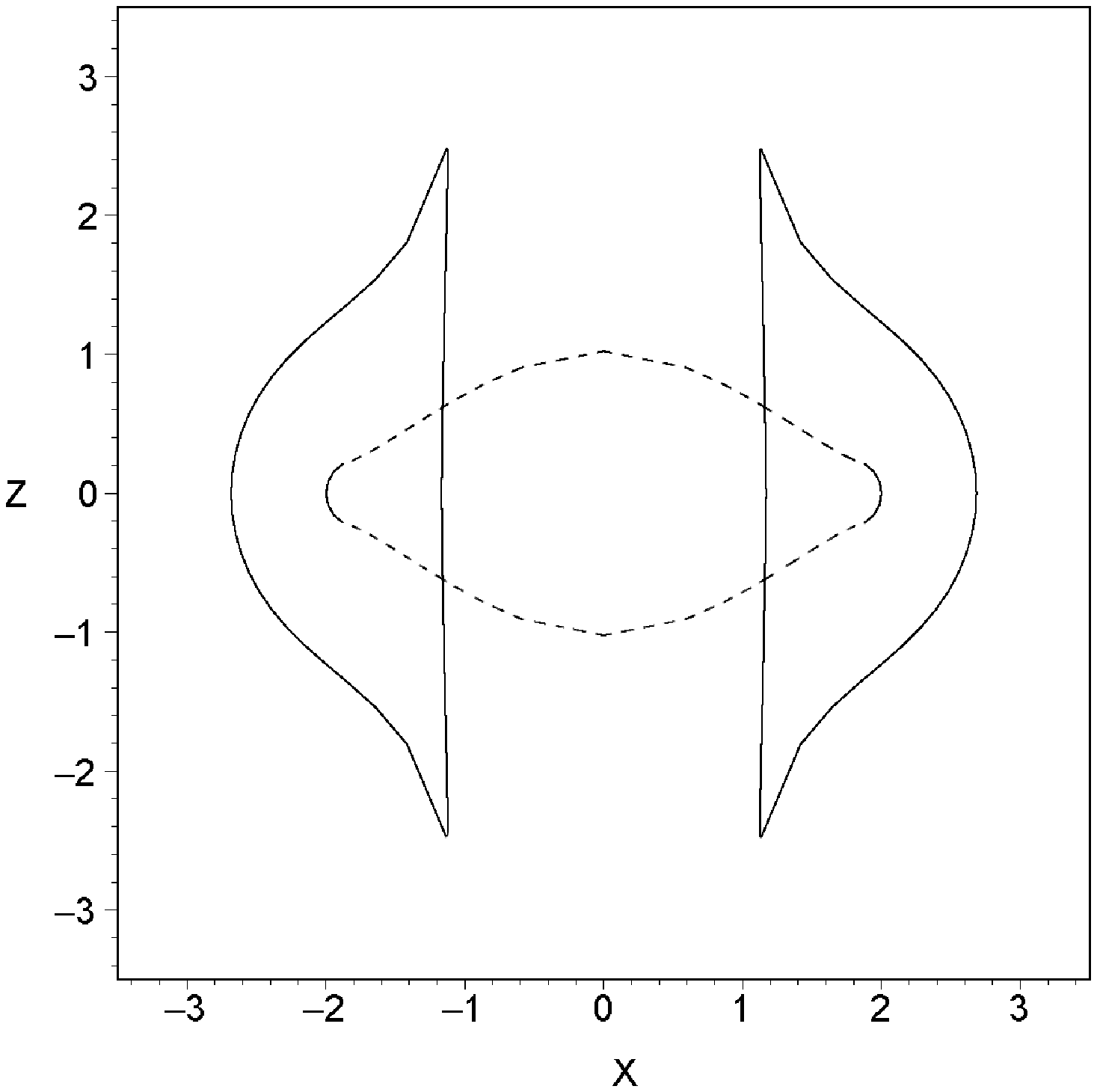}\\[.4cm]
\mbox{(c)} & \mbox{(d)}\\
\end{array}$
\end{center}
\caption{The projections on the $X-Z$ plane of the embedding diagrams of Fig. \ref{fig:4} are shown.
Dashed lines correspond to the Minkowskian part of the embedding of the outer horizon.
}
\label{fig:5}
\end{figure*}

The projections on the $X-Z$ plane of the embedding diagrams of Fig. \ref{fig:4} are shown in Fig. \ref{fig:5}.


\section{On the energy of the dyadoregion}

The total electromagnetic energy distributed in a stationary spacetime can be
determined by evaluating the conserved Killing integral (see e.g. \cite{vitagliano})
\begin{equation}
\label{Eemdef2}
E(\xi)=\int_{\Sigma} T_{\mu\nu}^{(\rm em)}\xi^\mu d\Sigma^\nu\ ,
\end{equation}
where $\xi=\partial_t$ is the timelike Killing vector, $T_{\mu\nu}^{(\rm em)}$
is the electromagnetic energy-momentum tensor of the source, $d\Sigma^\nu=n^\nu
d\Sigma$ is the surface element vector with $n$ the unit timelike normal to the
smooth compact spacelike hypersurface $\Sigma$. The integration is meant to be
performed through the whole spacetime occupied by the electromagnetic field,
i.e. by allowing $\Sigma$ to extend up to the spatial infinity. Evaluating the
electromagnetic energy stored inside a finite region with boundary $r=\,$ const
of spacetime would require
instead the introduction of the concept of \lq\lq quasilocal energy.'' However,
it is interesting to compare the results of the quasilocal treatment with the
expression of the electromagnetic energy contained in the portion of spacetime
with boundary $r=\,$ const obtained simply by
truncating the integration over $r$ at a given $R$ in Eq. (\ref{Eemdef2})
\begin{eqnarray}
\label{Eemxi}
E(\xi)_{(r_+,R)}&=&\int_{r_+}^{R} \int_{0}^{\pi} \int_{0}^{2\pi} \mathcal{E}(\xi)\sqrt{h_n} dr d\theta d\phi\nonumber\\
&=&\frac{Q^2}{4r_+}\left(1-\frac{r_+}{R}\right)+\frac{Q^2}{4r_+}\left[\left(1+\frac{a^2}{r_+^2}\right)\right.\nonumber\\
&&\left.\times\frac{\arctan(a/r_+)}{a/r_+}-\frac{r_+}{R}\left(1+\frac{a^2}{R^2}\right)\right.\nonumber\\
&&\left.\times\frac{\arctan(a/R)}{a/R}\right]\ ,
\end{eqnarray}
where
\begin{eqnarray}
\mathcal{E}(\xi)&=&T_{\mu\nu}^{\rm (em)}\xi^\mu n^\nu\nonumber\\
&=&\frac{Q^2}{8\pi\Sigma^{5/2}}\sqrt{\Delta}\frac{r^2-a^2\cos^2\theta+2a^2}{\sqrt{(r^2+a^2)^2-\Delta
a^2\sin^2\theta}}\
\end{eqnarray}
can be interpreted as the electromagnetic energy density, $n$ is the unit
normal to the time coordinate hypersurfaces and $h_n=(\Sigma/\Delta)[(r^2+a^2)^2-\Delta a^2\sin^2\theta]\sin^2\theta$ is the determinant of the induced metric. It is interesting to note that the same
results can be obtained by using the theory of pseudotensors \cite{virbhadra}
(see Appendix B). In the limit of vanishing rotation parameter Eq. (\ref{Eemxi}) becomes
\begin{equation}
\label{EemxiRN}
E(\xi)_{(r_+,R)}=\frac{Q^2}{2r_+}\left(1-\frac{r_+}{R}\right)\ ,
\end{equation}
which is just the expression for the electromagnetic energy obtained by
Vitagliano and Ruffini \cite{vitagliano} for the Reissner-Nordstr\"om geometry.
Eq. (\ref{Eemdef2}) can be actually considered as a possible quasilocal
definition of energy \cite{szabados}, although it strongly depends on the
existence of certain spacetime symmetries, i.e. the existence of a timelike
Killing vector, which characterizes stationary spacetimes.
In addition, we can see that since the current $J^\mu(\xi)=T^\mu_{\rm
(em)}{}_\nu\xi^\nu$ is a conserved vector, the resulting energy does not depend
on the chosen cut through spacetime.
In contrast, in any given spacetime one can always introduce a physically motivated congruence of observers $U$ measuring the energy irrespective of spacetime symmetries.
But the current $J^\mu(U)=T^\mu_{\rm (em)}{}_\nu U^\nu$ is not a conserved vector in general.
Therefore, in this case the energy has an observer dependent meaning; in addition, the results of the measurement could be different for different cuts through spacetime.

Such an approach consists of using the definition \cite{katz1,katz2}
\begin{equation}
\label{Edef}
E_{\Sigma}(U)=\int_{\Sigma} T^{(\rm em)}_{\mu\nu}U^\mu d\Sigma^\nu\ ,
\end{equation}
where $\Sigma$ is now a bounded hypersurface containing only a finite portion
of spacetime, and $U$ is the 4-velocity of the observer measuring the energy.
In general the flux integral of the current $J^\mu(U)=T^\mu_{\rm (em)}{}_\nu
U^\nu$ depends on the hypersurface, because this is not connected with the
spacetime symmetries. In particular, the vector field $U$ can be chosen to be
the unit timelike normal $n$ of $\Sigma$.  Therefore, generally we may always
evaluate $E_{\Sigma}(U)$ with respect to any preferred observer $U$, but should
not expect to get an answer independent of the chosen cut. In the case of
axially symmetric spacetimes in practice there is normally a good time
coordinate such as Boyer-Lindquist in Kerr and cuts are chosen to be at
constant time.  The current $J^\mu(U)$ will be conserved both for static
observers and ZAMOs (Zero Angular Momentum Observers), since their 4-velocities
are aligned with Killing vectors.

Due to the spacetime symmetries it is indeed quite natural to consider in the
Kerr-Newman spacetime two families of observers which are described by two
geometrically motivated congruences of curves: 1) static observers, at rest at
a given point in the spacetime, whose 4-velocity
$m=1/\sqrt{g_{tt}}\,\partial_t$ is aligned with the Killing temporal direction;
2) ZAMOs, a family of locally nonrotating observers with 4-velocity
$n=N^{-1}(\partial_t-N^\phi\partial_\phi)$, where $N=(-g^{tt})^{-1/2}$ and
$N^{\phi}=g_{t\phi}/g_{\phi\phi}$ are the lapse and shift functions
respectively,   characterized as that normalized linear combination of the two
given Killing vectors which is orthogonal to $\partial_\phi$ and
future-pointing, and it is the unit normal to the time coordinate
hypersurfaces.  Since the static observers do not exist inside the ergosphere,
the ZAMOs seem to be the best candidates to construct the energy (\ref{Edef}).
However, their 4-velocity diverges at the horizon, since the lapse function
goes to zero there.

In order to obtain a finite energy at the horizon one can then chose a family
of infalling observers as the Painlev\'e-Gullstrand observers, which move
radially with respect to ZAMOs and form a congruence of geodesic and
irrotational orbits, whose 4-velocity is given by
$U_{PG}=N^{-1}(n-\sqrt{1-N^2}e_{\hat r})$.  Since they do not follow the
spacetime symmetries the current $J^\mu(U_{PG})=T^\mu_{\rm (em)}{}_\nu
U_{PG}^\nu$ is not conserved, so the corresponding energy $E_{\Sigma}(U_{PG})$
depends on the hypersurface.  The result is that the expression (\ref{Eemxi})
of the electromagnetic energy contained in the dyadoregion constructed by
means of the (not normalized) timelike Killing vector agrees with the
electromagnetic energy assessed by the Painlev\'e-Gullstrand geodesic family of
infalling observers through the $T=\,$ const cut of the Kerr-Newman spacetime,
where $T$ denotes the Painlev\'e-Gullstrand time coordinate, i.e.
\begin{eqnarray}
\hspace{-.1cm}E_\Sigma(\xi)&\equiv&\hspace{-.3cm}
\underbrace{\int_{\Sigma} T_{\mu\nu}^{(\rm em)}\xi^\mu d\Sigma^\nu}_{{\scriptsize \mbox{BL coordinates, Killing vector, $t=\,$ const cut}}}\nonumber\\
&=&\hspace{-.3cm}
\underbrace{\int_{\Sigma} T_{\mu\nu}^{(\rm em)}{\mathcal{N}}^\mu d\Sigma^\nu}_{{\scriptsize \mbox{PG coordinates, PG 4-velocity, $T=\,$ const cut}}}
\equiv E_\Sigma(\mathcal{N}),\nonumber\\
\end{eqnarray}
with $\mathcal{N}$ the timelike normal to the chosen cut.  Details can be found in Appendix B.

From Eq. (\ref{Eemxi}), a rough estimate of the electromagnetic energy stored inside the \lq\lq dyadoregion'' turns out to be given by $E(\xi)_{(r_+,R)}\approx5.5\times10^{-3}$ cm $\approx6.7\times10^{46}$ ergs by assuming $R=2r_+$ with the same parameters as in Fig. \ref{fig:2} (d), and $E(\xi)_{(r_+,R)}\approx1.9\times10^{-2}$ cm $\approx2.3\times10^{47}$ ergs if $R=3r_+$ with the same choice of parameters as in Fig. \ref{fig:2} (a).
We note that an exact analytic expression for the electromagnetic energy can also be obtained by taking the actual shape $r=r^d_\pm$ given by Eq. (\ref{dyadosurf}) instead of the approximate expression $r=R=\,$ const in the evaluation of the integral (\ref{Eemxi}). However, this only complicates matters by introducing a nontrivial dependence on the polar angle $\theta$ which makes the integration procedure more involved, even if it can be analytically performed (not shown here for the sake of brevity). Furthermore, the numerical values of the energy corresponding to the above choice of parameters agree with previous estimates.

It is interesting to compare the electromagnetic energy (\ref{Eemxi}) of an
extreme Kerr-Newman black hole contained in the portion of spacetime with
boundary $R=\,$ const and that of a Reissner-Nordstr\"om black hole
(\ref{EemxiRN}) with the same total mass and charge in the limit of small charge to mass ratio.
In this limit we have
\begin{eqnarray}
E_{\rm RN} &\simeq& \frac{Q^2}{4 M}\left( 1-\frac{2M}{R} \right)\ , \nonumber\\
E_{\rm KN} &\simeq& \frac{Q^2}{4 M}\left( 1-\frac{2M}{R} \right)\\
&&+\frac{Q^2}{4 M}\left[\frac{\pi}{2}+\frac{M}{R}-\left(1+\frac{M^2}{R^2}\right)\arctan(M/R) \right]\ \nonumber.
\end{eqnarray}
A comparison between energies is meaningful only at infinity, where the radial coordinates of a Kerr-Newman and a Reissner-Nordstr\"om geometry can be identified (both with an ordinary radial coordinate in flat space).
For $R\to\infty$ we thus have
\begin{equation}
E_{\rm KN}-E_{\rm RN} \to \frac{Q^2}{4 M} \frac{\pi}{2}>0\ .
\end{equation}

\section{Conclusions}

Vacuum polarization processes can occur in the field of a Kerr-Newman black hole inside a region we have called dyadotorus, whose properties have been investigated here.
Such a region has an invariant character, i.e. its existence does not depend on the observer measuring the electromagnetic field: therefore, it is a true physical region.

Some pictorial representations of the boundary surface similar to those commonly used in the literature have been shown employing Cartesian-like coordinates (i.e. ordinary spherical coordinates built up simply using the Boyer-Lindquist radial and angular coordinates) as well as Kerr-Schild coordinates.
The dyadotorus has been also shown on the corresponding embedding diagram, which gives the correct geometry allowing to visualize the spacetime curvature.

We have then estimated the electromagnetic energy contained in the dyadotorus by using three different approaches, which give rise to the same final expression for the energy.
The first one follows the standard approach consisting of using the (not normalized) timelike Killing vector through the Boyer-Lindquist constant time cut of the Kerr-Newman spacetime (see e.g. \cite{vitagliano}), the second one follows a recent observer dependent definition by Katz, Lynden-Bell and Bi{\v c}\'ak \cite{katz1,katz2} for axially
symmetric asymptotically flat spacetimes, for which we have used the Painlev\'e-Gullstrand geodesic family of infalling observers through the Painlev\'e-Gullstrand constant time cut, and the last one adopts the pseudotensor theory (see e.g. \cite{virbhadra}).
We have found by rough estimates that the extreme Kerr-Newman black hole leads to larger values of the electromagnetic energy as compared with a Reissner-Nordstr\"om black hole with the same total mass and charge.

It is appropriate to recall that the release of energy via the electron-positron pairs in the dyadotorus is the most powerful way to extract energy from black holes and in all sense corresponds to a new form of energy: the ``blackholic'' energy \cite{ruffinikerr}.
This is a new form of energy different from the traditional ones known in astrophysics. The thermonuclear energy has been recognized to be energy source of main sequence stars lasting for $10^9$ years \cite{bethe}, the gravitational energy released by accretion processes in neutron stars and black holes has explained the energy observed in binary X-ray sources on time scales $10^6-10^8$ years \cite{giacconi}.
The ``blackholic'' energy appears to be energy source for the most transient and most energetic events in the universe, the GRBs \cite{ruffinikerr}.

\begin{acknowledgments}
The authors thank Dr. C. L. Bianco for the technical assistance plotting Fig. \ref{fig:2bis}.
\end{acknowledgments}

\appendix

\section{Newman-Penrose quantities and invariant definition of the dyadotorus}

The existence of the dyadotorus has an invariant character.
This fact appears more evident if the electric and magnetic field strengths are expressed in terms of the electromagnetic invariants.
Let us adopt here the metric signature $(+,-,-,-)$ in order to use the Newman-Penrose formalism in its original form and then easily get the necessary physical quantities \cite{chandra,bccr}.
The Kerr-Newman metric is thus given by
\begin{eqnarray}
\label{KNmetricNP}
\hspace{-1cm}ds^2\hspace{-.1cm}&=&\hspace{-.1cm}\left( 1- \frac{2Mr-Q^2}{\Sigma} \right) dt^2\nonumber\\
&&\hspace{-.1cm} +\frac{ 2a\sin^2\theta }{\Sigma}\left(2Mr-Q^2\right) dt d\phi-\frac{\Sigma}{\Delta} dr^2-\Sigma d\theta^2\nonumber\\
&&\hspace{-.1cm}- \left[r^2+a^2+\frac{a^2\sin^2\theta}{\Sigma}(2Mr-Q^2)\right]\sin^2\theta d\phi^2,
\end{eqnarray}
with associated electromagnetic field
\begin{eqnarray}
\hspace{-1cm}F&=& \frac{Q}{\Sigma^2}(r^2-a^2\cos^2\theta)dr \wedge [dt -a \sin^2 \theta d\phi]\nonumber \\
&& +2\frac{Q}{\Sigma^2}ar \sin \theta \cos \theta d\theta \wedge [(r^2+a^2)d\phi - a d\phi].
\end{eqnarray}
Introduce the standard Kinnersley principal tetrad \cite{kinnersley}
\begin{eqnarray}
\label{kinn}
l^{\mu}&=&\frac{1}{\Delta}[r^2+a^2,\Delta,0,a]\ , \nonumber\\
n^{\mu}&=&\frac{1}{2\Sigma}[r^2+a^2,-\Delta,0,a]\ , \nonumber\\
m^{\mu}&=&\frac{1}{\sqrt{2}(r+ia\cos\theta)}\,\left[{ia}\,{\sin\theta},0,1,\frac{i}{\sin\theta}\right]\ ,
\end{eqnarray}
which gives nonvanishing spin coefficients
\begin{eqnarray}
\rho&=&-\frac{1}{r-ia\cos\theta}\ , \qquad
\tau=-\frac{ia}{\sqrt{2}}\rho\rho^{*}\sin\theta\ , \nonumber\\
\beta&=&-\frac{\rho^{*}}{2\sqrt{2}}\cot\theta\ , \qquad
\pi=\frac{ia}{\sqrt{2}}\rho^2\sin\theta\ , \nonumber\\
\mu&=&\frac12\rho^2\rho^{*}\Delta\ , \qquad
\gamma=\mu+\frac12\rho\rho^{*}(r-M)\ , \nonumber\\
\alpha&=&\pi-\beta^{*}\ ,
\end{eqnarray}
and the only nonvanishing Weyl scalar
\begin{equation}
\psi_2=M\rho^3+Q^2\rho^*\rho^3\ ,
\end{equation}
showing clearly the Petrov type D nature of the Kerr-Newman spacetime, whereas the Maxwell scalars are
\begin{equation}
\label{maxscal}
\phi_0=\phi_2=0\ , \qquad \phi_1=\frac{Q}{2}\rho^2\ .
\end{equation}

The electromagnetic invariants are given by
\begin{eqnarray}
\label{invar}
{\mathcal F}&\equiv&\frac14F_{\mu\nu}F^{\mu\nu}=\frac12({\bf B}^2-{\bf E}^2)=2{\rm Re}(\phi_0\phi_2-\phi_1^2)\ , \nonumber\\
{\mathcal G}&\equiv&\frac14F_{\mu\nu}{}^*F^{\mu\nu}={\bf E}\cdot {\bf B}=-2{\rm Im}(\phi_0\phi_2-\phi_1^2)\ ,
\end{eqnarray}
where ${\bf E}$ and ${\bf B}$ are the electric and magnetic fields.
Requiring parallel electric and magnetic fields \cite{damruff} as measured by the Carter observer \cite{carter}, the previous relations become
\begin{eqnarray}
\label{invar2}
|{\bf B}|^2-|{\bf E}|^2=-4{\rm Re}(\phi_1^2)\ , \quad
|{\bf E}|\,|{\bf B}|=2{\rm Im}(\phi_1^2),
\end{eqnarray}
taking into account Eq. (\ref{maxscal}).
This system  can then be easily solved for the magnitudes of ${\bf E}$ and ${\bf B}$ in the Kerr-Newman background, which turn out to be given by
\begin{eqnarray}
|{\bf E}|=\left\vert\frac{Q}{\Sigma^2}(r^2-a^2\cos^2\theta)\right\vert, \,\,
|{\bf B}|=\left\vert2\frac{Q}{\Sigma^2}ar\cos\theta\right\vert,
\end{eqnarray}
which coincide with those of Eq. (\ref{EBsols}).
We have thus recovered the results by Damour and Ruffini \cite{damruff}, but using a different faster derivation using the Newman-Penrose formalism.

Finally, the Schwinger formula for the rate of pair creation per unit four-volume in terms of the electromagnetic invariants (\ref{invar}) is given by \cite{schwinger}
\begin{eqnarray}
2{\rm Im}\mathcal L&=&\frac{e^2|{\mathcal G}|}{4\pi^2\hbar^2}\sum_{n=1}^\infty\frac1{n}\coth\left\{n\pi\left[\frac{({\mathcal F}^2+{\mathcal G}^2)^{1/2}+{\mathcal F}}{({\mathcal F}^2+{\mathcal G}^2)^{1/2}-{\mathcal F}}\right]^{1/2}\right\}\nonumber\\
&&\times
e^{-n\pi E_c/[({\mathcal F}^2+{\mathcal G}^2)^{1/2}-{\mathcal F}]^{1/2}}\ .
\end{eqnarray}
After introducing the Carter frame (\ref{Ucarter})--(\ref{carterframe}) with respect to which electric and magnetic fields are parallel, the previous formula reduces to Eq. (\ref{schwingerKN}), since
\begin{eqnarray}
\hspace{-1cm}&&[({\mathcal F}^2+{\mathcal G}^2)^{1/2}+{\mathcal F}]^{1/2}=|{\bf B}|\ , \nonumber\\
\hspace{-1cm}&&[({\mathcal F}^2+{\mathcal G}^2)^{1/2}-{\mathcal F}]^{1/2}=|{\bf E}|, \quad
|{\mathcal G}|=|{\bf E}|\,|{\bf B}|\ .
\end{eqnarray}

\section{Electromagnetic energy using Painlev\'e-Gullstrand observers and pseudotensor theory}

In order to evaluate the energy $E_{\Sigma}(U_{PG})$ it is useful to transform the Kerr-Newman metric (\ref{KNmetric}) from Boyer-Lindquist coordinates $(t,r,\theta,\phi)$ to Painlev\'e-Gullstrand coordinates $(T,R,\Theta,\Phi)$ \cite{doran,cook}, which are related by the transformation
\begin{eqnarray}
T&=& t-\int^r f(r) dr\ , \qquad
R=r\ , \qquad
\Theta=\theta\ , \nonumber\\
\Phi &=& \phi -\int^r \frac{a}{r^2+a^2} f(r) dr\ ,
\end{eqnarray}
where
\begin{equation}
f(r)=-\frac{\sqrt{(2Mr-Q^2)(r^2+a^2)}}{\Delta}\ .
\end{equation}
Let us notice that $r$ and $R$ are identified. This is also true for their differential $dr=dR$ but it is no more true for the associated differentiations $\partial_r \not = \partial_R$. Hereafter we will always use $r$ in place of $R$, except for the differentiation operations.
In differential form,  this transformation writes as
\begin{eqnarray}
\label{forms}
dT&=& dt- f(r) \, dr\ ,\quad dR=dr\ ,\quad d\Theta=d\theta\ , \nonumber\\
d\Phi &=& d\phi -\frac{a}{r^2+a^2} f(r)\, dr\ .
\end{eqnarray}
Finally, the Kerr-Newman metric in the Painlev\'e-Gullstrand coordinates is given by
\begin{eqnarray}
\label{PGkn}
ds^2&=&-\left(1-\frac{2Mr-Q^2}{\Sigma}\right) dT^2+ 2 \sqrt{\frac{2Mr-Q^2 }{r^2+a^2}} dT dr
\nonumber \\
& & -\frac{2a(2Mr-Q^2) }{\Sigma}\sin^2\theta dTd\Phi \nonumber \\
&& +\sin^2 \theta \left[r^2+a^2+\frac{a^2(2Mr-Q^2)}{\Sigma}\sin^2\theta \right] d\Phi^2\nonumber \\
& &  -2a\sin^2\theta\sqrt{\frac{2Mr-Q^2}{r^2+a^2}} dr d\Phi
\nonumber \\
&& +\frac{\Sigma}{r^2+a^2} dr^2 + \Sigma d\theta^2\ ,
\end{eqnarray}
with associated electromagnetic field
\begin{eqnarray}
\label{FknPG}
&&F= \frac{Q}{\Sigma^2}(r^2-a^2\cos^2\theta)dr \wedge [dT -a \sin^2 \theta d\Phi]\nonumber \\
&& +2\frac{Q}{\Sigma^2}ar \sin \theta \cos \theta d\theta \wedge [(r^2+a^2)d\Phi - a dT]\ ,
\end{eqnarray}
which has the same form as (\ref{Fkn}) with $dt\to dT$ and $d\phi\to d\Phi$.

The limit of vanishing rotation parameter $a=0$ of the previous equations (\ref{PGkn})--(\ref{FknPG}) gives rise to the Reissner-Nordstr\"om solution in Painlev\'e-Gullstrand coordinates
\begin{eqnarray}
ds^2&=&-\left(1-\frac{2M}{r}+\frac{Q^2}{r^2}\right)dT^2 + 2\frac{\sqrt{2Mr-Q^2}}{r} dT dr\nonumber \\
& & +dr^2+r^2(d\theta^2 +\sin^2 \theta  d\phi^2)\ , \nonumber\\
F&=& \frac{Q}{r^2}dr \wedge dT\ .
\end{eqnarray}

In the Painlev\'e-Gullstrand coordinates the slicing observers ($T-$slicing hereafter) have 4-velocity
\begin{equation}
\label{PG4vel}
\mathcal{N}=\partial_T -\frac{\sqrt{(2Mr-Q^2)(r^2+a^2)}}{\Sigma}\partial_R\
\end{equation}
and associated 1-form $\mathcal{N}^\flat=-dT$.
This family of $T-$slicing-adapted observers  does not coincide with  the $t-$slicing-adapted observers in Boyer-Lindquist coordinates  once the coordinate transformation is performed.
In fact, when expressed in Boyer-Lindquist coordinates the $T-$slicing-adapted observers move with respect to the
$t-$slicing-adapted observers in the radial direction, as already pointed out in Section V.

We are now ready to evaluate the energy (\ref{Edef}) through a $T=\,$ const hypersurface as measured by Painlev\'e-Gullstrand observers with 4-velocity (\ref{PG4vel}). The energy density turns out to be
\begin{equation}
\mathcal{E}(\mathcal{N})=T_{\mu\nu}^{\rm (em)}\mathcal{N}^\mu\mathcal{N}^\nu=\frac{Q^2}{8\pi\Sigma^3}(r^2-a^2\cos^2\theta+2a^2)\ ,
\end{equation}
where $T_{\mu\nu}^{\rm (em)}$ is the Kerr-Newman electromagnetic energy-momentum tensor expressed in Painlev\'e-Gullstrand coordinates.
Let us assume that the boundary $S$ of $\Sigma$ be the 2-surface $r=R=\,$ const for simplicity.
Therefore the energy (\ref{Edef}) turns out to be given by
\begin{eqnarray}
\label{PGqle}
E(\mathcal{N})_{(r_+,R)}&=&2\pi\int_{r_+}^{R}\int_0^\pi\mathcal{E}(\mathcal{N})\sqrt{h_{\mathcal{N}}}drd\theta\nonumber \\
&=&-\frac{Q^2}{4a}\left[\frac{a}r+\frac{r^2+a^2}{r^2}\arctan\frac{a}{r}-\frac{\pi}{2}\right]^{R}_{r_+}\nonumber\\
&=&\frac{Q^2}{4r_+}-\frac{Q^2}{4R}+\frac14\frac{Q^2}{ar_+^2}(r_+^2+a^2)\arctan\frac{a}{r_+}\nonumber \\
& & -\frac14\frac{Q^2}{aR^2}(R^2+a^2)\arctan\frac{a}{R}\ ,
\end{eqnarray}
where $h_{\mathcal{N}}=\Sigma^2\sin^2\theta$ is the determinant of the induced metric.
The total electromagnetic energy contained in the whole spacetime is obtained by taking the limit $R\to\infty$ in the previous equation
\begin{equation}
E(\mathcal{N})_{(r_+,\infty)}=\frac{Q^2}{4r_+}+\frac14\frac{Q^2}{ar_+^2}(r_+^2+a^2)\arctan\frac{a}{r_+}\ ,
\end{equation}
which in the limiting case $a=0$ reduces to
\begin{equation}
E^{\rm RN}(\mathcal{N})_{(r_+,\infty)}=\frac{Q^2}{2r_+}\ .
\end{equation}

It is interesting to note that the same result (\ref{PGqle}) for the energy assessed by Painlev\'e-Gullstrand observer is achieved simply by using the Killing vector $\xi=\partial_T$, since $\mathcal{E}(\mathcal{N})=\mathcal{E}(\xi)$.
But it is quite surprising that the same result is again obtained by taking a $t=\,$ const hypersurface in Boyer-Linquist coordinates with unit normal the ZAMO 4-velocity $n$ with respect to \lq\lq Killing observers'' $\xi=\partial_t$ (see Eq. (\ref{Eemxi})), since
\begin{equation}
\mathcal{E}(\mathcal{N})\sqrt{h_{\mathcal{N}}}=\mathcal{E}\sqrt{h_n}=\frac{Q^2}{8\pi\Sigma^2}(r^2-a^2\cos^2\theta+2a^2)\sin\theta\ .
\end{equation}

For completeness we list here similar results presented in Ref. \cite{virbhadra} by using the standard definition of symmetric energy-momentum pseudotensor as given by Landau and Lifshitz \cite{landau} (LL), although we stress that the physical interpretation of these quantities are controversial in the
literature, due to their strict relation with specific coordinate sets.
This fact is clearly not in the spirit of general relativity.
The LL prescription for the pseudotensor is given by $16\pi L^{\alpha\beta}=\lambda^{\alpha\beta\gamma\delta}{}_{,\gamma\delta}$, where comma denotes partial derivative and $\lambda^{\alpha\beta\gamma\delta}=-g\left( g^{\alpha\beta}g^{\gamma\delta}-g^{\alpha\gamma}g^{\beta\delta}\right)$.
The conservation law $L^{\alpha\beta}{}_{,\beta}=0$ implies that the total energy is given by $E=\int\int\int L^{00}dx^1dx^2dx^3$.
By computing the pseudotensor in the quasi-Cartesian Kerr-Schild coordinates previously introduced in Eq. (\ref{kerrschild}) and requiring the integration to be performed on a Boyer-Lindquist $r=R=\,$ const surface, one obtains the result $E=M-K$, where $K$ is just the r.h.s. of Eq. (\ref{PGqle}).
Note that in Ref. \cite{virbhadra} the same result is obtained using plenty other different energy-momentum pseudotensors.

\end{document}